\documentclass[
  aps,
  pre,
  twocolumn,
  groupedaddress,
  floatfix,
  longbibliography,
  superscriptaddress,
  10pt
]{revtex4-2}

\usepackage{amsmath}
\usepackage{amssymb}
\usepackage{amsfonts}
\usepackage{mathrsfs}
\usepackage{bm}
\usepackage{bbold}
\usepackage{braket}
\usepackage{slashed}
\usepackage{cancel}
\usepackage[dvipsnames]{xcolor}
\usepackage{graphicx}
\usepackage{epsfig}
\usepackage{subfigure}
\usepackage{multirow}

\usepackage{feynmf}

\usepackage{times}

\usepackage[dvipsnames,svgnames]{xcolor}

\usepackage{hyperref}

\hypersetup{
  colorlinks=true,
  linkcolor=Blue,
  citecolor=Blue,
  urlcolor=Blue
}

\usepackage{accents}
\usepackage{ulem}
\usepackage{simpler-wick}
\usepackage{orcidlink}

\DeclareUnicodeCharacter{039B}{\ensuremath{\Lambda}}
\usepackage{accents}
\usepackage{ulem}
\usepackage{simpler-wick}
\usepackage{orcidlink}

\newcommand{\bea}{\begin{eqnarray}}
\newcommand{\eea}{\end{eqnarray}}
\newcommand{\be}{\begin{equation}}
\newcommand{\ee}{\end{equation}}
\newcommand{\nn}{\nonumber}
\newcommand{\ii}{\mathrm{i}}
\newcommand{\sumint}{\mathop{\vcenter{\hbox{\ooalign{\hfil$\displaystyle\sum$\hfil\cr\hfil$\displaystyle\int$\hfil\cr}}}}}

\begin{document}

\title{Finite-Size Effects on Symmetry Restoration in a Rotating Scalar System with Spatially Dependent Thermal Self-Energy}

\author{Jorge David Casta\~no-Yepes~\orcidlink{0000-0002-8654-1304}}
\email{jorge.yepes@correounivalle.edu.co}
\affiliation{ Departamento de Física, Universidad del Valle, Ciudad Universitaria Meléndez,
Santiago de Cali 760032, Colombia}
\author{Larry A. Cerón-Suárez~\orcidlink{0009-0008-9548-0444}}
\affiliation{Programa de Física, Universidad del Valle, Ciudad Universitaria Meléndez,
Santiago de Cali 760032, Colombia}

\begin{abstract}
We investigate $\mathbb{Z}_2$ symmetry restoration in a real $\lambda\phi^4$ theory confined to a finite cylindrical region undergoing rigid rotation. The finite transverse extent is imposed through Dirichlet boundary conditions, resulting in a discrete Fourier-Bessel spectrum and an explicit radial dependence of the fluctuation propagator. Using the background-field method, we derive the one-loop effective potential while retaining the spatial structure induced by the finite geometry. At finite temperature, rotation modifies the thermal mode energies through angular-momentum-dependent shifts, and the coincident-point propagator generates a position-dependent thermal self-energy that is incorporated through ring resummation. We use the resulting spatially resolved effective potential to determine the symmetry-restoration temperature $T_c(\Omega,R)$ and examine its dependence on the angular velocity and transverse size. The results show that the finite transverse size of the system modifies the symmetry-restoration temperature, leading to a nontrivial dependence of $T_c(\Omega,R)$ on $R$. An additional scaling behavior emerges when the normalized critical temperature is expressed in terms of the boundary velocity $\Omega R$: systems with different transverse sizes exhibit the same relative modification of $T_c$ when their boundary velocities are equal. This scaling persists after the inclusion of the position-dependent thermal self-energy through ring resummation, indicating that the $\Omega R$ dependence of the rotational response is maintained in the interacting finite system.
\end{abstract}

\maketitle

\section{\bf Introduction}

Noncentral relativistic heavy-ion collisions generate strongly interacting matter carrying substantial orbital angular momentum. A fraction of this angular momentum can be transferred to the produced medium, giving rise to vortical motion whose magnitude and spatial structure depend on the collision geometry and on the subsequent evolution of the system~\cite{becattini2015vorticity,deng2016vorticity,ivanov2020vorticity}. The observation of global $\Lambda$ and $\overline{\Lambda}$ polarization provides experimental evidence for this vortical state~\cite{star2017polarization}, while its dependence on collision energy and centrality contains information about the space-time evolution and the relative contribution of different regions of the medium~\cite{PhysRevC.105.034907,ayala2026polarization}. These observations have motivated the incorporation of rotation into finite-temperature quantum field theory as a relevant ingredient in the description of strongly interacting matter~\cite{fukushima2019extreme}.

The medium produced in such collisions is also spatially finite. Its characteristic size depends on the collision geometry and evolves during the different stages of the collision, with femtoscopic measurements providing information on the spatial extent of the matter at freeze-out~\cite{PhysRevC.59.3324,PhysRevC.83.044910,PhysRevC.93.024905}. Finite size therefore affects more than the infrared sector of the theory: it modifies the spectrum of available excitations and can consequently influence the thermal properties of the system. In the presence of rotation, this finite-volume spectrum becomes particularly relevant because the rotational state couples to the angular momentum carried by the allowed modes. The combined effect of finite size and rotation thus requires a description in which the spatial spectrum is retained explicitly rather than replaced from the outset by a homogeneous thermodynamic approximation.

Rigid rotation in relativistic quantum field theory has been formulated in cylindrical geometries, with the rotation axis defining the symmetry axis of the system. Such formulations have been developed for scalar and fermionic fields and applied to the construction of propagators and to the study of thermal properties of rotating quantum systems~\cite{vilenkin1980rotation,siri2024bose,kawaguchi2025susceptibilities,salvio2026scalar,kuboniwa2026perturbation,hernandez2026scalar,castano2026dilepton}. The cylindrical geometry provides a natural spectral representation in which the angular dependence of the modes is characterized by their angular momentum around the rotation axis, while a finite radial boundary generates a discrete transverse spectrum. For scalar fields, this structure leads naturally to a Fourier-Bessel representation of the modes and to propagators with a nontrivial radial dependence~\cite{kuboniwa2026perturbation,hernandez2026scalar}. The finite radial extent is therefore not merely a geometric choice, but enters directly into the spectrum governing the thermal dynamics.

The cylindrical formulation is also closely connected with the causal structure of rigid rotation. For constant angular velocity, the tangential velocity increases with the distance from the rotation axis, reaching the speed of light at the corresponding light cylinder. A finite rotating system must therefore be contained within this causal region, which constrains the allowed combination of angular velocity and transverse size~\cite{vilenkin1980rotation,siri2024bose,kuboniwa2026perturbation,hernandez2026scalar}. This restriction defines the domain in which the rigidly rotating configuration can be consistently considered, but it does not by itself determine the thermal response of the system. Within the causal region, the finite radial boundary continues to affect the dynamics through the discrete transverse spectrum and its dependence on the system size. Recent studies of rotating effective QCD models have likewise emphasized the role of finite volume and the causal boundary in determining the thermodynamic behavior of rotating systems~\cite{PhysRevD.110.094053,4zrn-wgrg,knn8-sv3k}.

These features become especially relevant for thermal phase transitions. In a homogeneous system, symmetry restoration can be studied through an effective potential for a spatially uniform order parameter, with thermal fluctuations modifying the potential and potentially restoring symmetries broken at zero temperature~\cite{dolan1974symmetry,lebellac1996thermal,kapusta2006finite}. Rotation introduces an additional scale associated with the angular velocity and has been shown to modify chiral and deconfinement transition temperatures in several effective descriptions~\cite{gaspar2023chiral,hernandez2025vortical,siri2024bose,PhysRevD.103.094515,PhysRevD.110.094053,knn8-sv3k}. In particular, studies of rotating linear sigma models have examined the dependence of critical temperatures on the angular velocity and system size, while lattice simulations of gluodynamics have investigated the rotational modification of the confinement-deconfinement transition~\cite{PhysRevD.103.094515,PhysRevD.110.094053,knn8-sv3k}. Independently, finite-volume effects and fluctuations of the thermal environment can alter the location and structure of thermal transitions~\cite{PhysRevD.106.116019,PhysRevD.110.056014}. These developments motivate a more specific question for a rotating finite system: how is the rotational modification of a thermal transition affected by the finite spatial spectrum of the system?

A scalar $\lambda\phi^4$ theory provides a controlled setting in which this question can be addressed without the additional structure associated with gauge or fermionic degrees of freedom. In a finite rotating cylinder, the radial boundary determines a discrete set of transverse modes, while rotation modifies their thermal occupation according to their angular momentum. The thermal propagator therefore retains both the finite-volume spectrum and the rotational structure of the medium. The thermodynamic properties of rigidly rotating scalar fields in finite cylindrical systems have been studied in detail, including their dependence on the boundary velocity and the constraints imposed by the light cylinder~\cite{PhysRevD.108.085016}. Perturbative treatments of rotating scalar fields have subsequently established the corresponding spectral formulation~\cite{kuboniwa2026perturbation}, while recent work on the scalar propagator in a rotating thermal medium has examined its radial structure, its nonrotating limit, and its one-loop self-energy, including applications to the finite-temperature $\lambda\phi^4$ effective potential and ring resummation~\cite{hernandez2026scalar}. These results provide the field-theoretical basis for treating thermal fluctuations in a rotating finite system while retaining their spatial structure.

For a finite rotating system, the transverse boundary affects the thermal fluctuations through the discrete spectrum of radial modes. The resulting propagator retains an explicit dependence on the transverse coordinate, leading to a position-dependent thermal self-energy and a spatially resolved effective potential~\cite{kuboniwa2026perturbation,hernandez2026scalar}. The transition is therefore determined locally by the inhomogeneous effective potential and globally through its spatial average over the finite transverse domain. This distinction is essential when the finite geometry is retained in the thermal dynamics rather than introduced only as a restriction on the allowed modes.

The finite-size dependence must be distinguished from the causal restriction imposed by rigid rotation. The condition $\Omega R<1$ constrains the admissible rotational configurations, whereas the discrete transverse spectrum determines the thermal response within this domain~\cite{PhysRevD.108.085016,kuboniwa2026perturbation}. Varying $R$ changes the radial eigenvalues and their spacing, and consequently modifies the thermal fluctuations entering the effective potential. The problem therefore requires retaining the finite-volume spectral structure throughout the calculation rather than treating the boundary solely as a kinematic constraint.

In this work, we investigate the $\mathbb{Z}_2$ symmetry-restoration transition of a real $\lambda\phi^4$ field confined to a finite rotating cylinder with Dirichlet boundary conditions. We retain the radial dependence of the thermal propagator and incorporate the resulting position-dependent thermal self-energy through ring resummation. This allows us to determine the spatially resolved effective potential and, after spatial averaging over the finite transverse domain, the critical temperature $T_c(\Omega,R)$. We then examine its dependence on the angular velocity and transverse size, distinguishing the dynamical finite-size effects from the causal restriction imposed by rigid rotation.

The manuscript is organized as follows. Section~\ref{sec:theory} presents the formulation of the finite rotating scalar system, including the field modes, scalar propagator, background-field effective potential, finite-temperature contributions, and ring resummation. Section~\ref{sec:results} presents the resulting critical temperature and its dependence on the angular velocity and transverse size. Section~\ref{sec:conclusions} summarizes the results. The technical details of the rotating geometry are derived in Appendix~\ref{sec:rotating_geometry}, while the scalar propagator and its Fourier-Bessel representation are developed in Appendix~\ref{app:scalar_propagator}. The functional derivation of the one-loop effective action and the finite-temperature contributions are presented in Appendix~\ref{app:functional-proofs}. Appendix~\ref{app:ring-resummation} develops the functional formulation of the ring resummation in the presence of transverse translational-symmetry breaking. Finally, Appendix~\ref{app:high_temp_expansion} derives the high-temperature approximation of the spatially averaged effective potential, including the analytical reduction of the longitudinal momentum and auxiliary resummation-parameter integrations.
\section{Rotating Scalar Field in a Finite Cylindrical System}\label{sec:theory}
We consider a real scalar $\lambda\Phi^4$ theory confined to a finite cylindrical region of radius $R$ undergoing rigid rotation around the $z$ axis with angular velocity $\Omega$. The model provides a minimal field-theoretical framework for investigating how finite size and rotation modify the structure of a symmetry-breaking potential at finite temperature. Although it does not contain the microscopic degrees of freedom of QCD, the $\lambda\Phi^4$ theory captures the essential mechanism of spontaneous symmetry breaking and its restoration, making it a useful starting point for studying these effects before considering more complete effective descriptions of QCD matter, such as the linear sigma model. This connection is particularly relevant in the context of ultrarelativistic heavy-ion collisions, where the strongly interacting matter produced in non-central collisions occupies a finite region and carries substantial angular momentum.

The scalar field is described by the Lagrangian density
\bea
\mathcal{L}&=&\frac{1}{2}\partial_\mu\Phi\partial^\mu\Phi-\frac{1}{2}m^2\Phi^2-\frac{\lambda}{4!}\Phi^4,
\label{eq:scalar_lagrangian_main}
\eea
where $m^2<0$ and $\lambda>0$, so that the classical potential exhibits spontaneous symmetry breaking. The finite cylindrical geometry provides an idealized description of the spatial extent of the system, while the rigid rotation introduces the angular velocity $\Omega$ as a control parameter. To describe the system in the rotating frame, we introduce cylindrical coordinates $(t,r,\phi,z)$, where $t$ denotes the time coordinate, $r$ is the radial distance from the rotation axis, $\phi$ is the azimuthal angle, and $z$ is the coordinate along the rotation axis. The corresponding line element, derived in Appendix~\ref{sec:rotating_geometry}, is
\bea
ds^2&=&(1-\Omega^2r^2)dt^2-2\Omega r^2dt\,d\phi-dr^2-r^2d\phi^2-dz^2.\nn\\
\label{eq:rotating_line_element_main}
\eea
where $ds^2$ denotes the invariant spacetime interval, and the radial coordinate is restricted to $0\leq r\leq R$. The rotating frame is physically well defined within the causal region $\Omega R<1$. At the cylindrical boundary, the scalar fluctuations satisfy Dirichlet boundary conditions,
\bea
\Phi(t,R,\phi,z)&=&0.
\label{eq:dirichlet_boundary_main}
\eea

The boundary condition fixes the allowed transverse momenta, which are quantized by the radial eigenvalue problem. As discussed in Appendix~\ref{app:scalar_propagator}, the corresponding discrete spectrum is
\bea
k_{\perp,l,h}&=&\frac{\gamma_{l,h}}{R},
\label{eq:transverse_momentum_main}
\eea
where $k_{\perp,l,h}$ denotes the momentum transverse to the rotation axis, $l\in\mathbb{Z}$ labels the azimuthal modes, $h=1,2,\ldots$ labels the radial modes, and $\gamma_{l,h}$ is the $h$th positive zero of the Bessel function $J_l$. The resulting Fourier-Bessel representation of the scalar propagator, which incorporates the spatial structure imposed by the finite boundary and the rotational dependence of the modes, is derived in Appendix~\ref{app:scalar_propagator}.

To construct the finite-temperature field theory, we use this propagator as the starting point for the thermal formulation. The loss of translational invariance in the transverse plane, induced by the finite cylindrical boundary, implies that the propagator is not translationally invariant in those directions. In particular, its coincidence limit retains an explicit dependence on the radial coordinate through the Fourier-Bessel modes. This spatial dependence will also be inherited by the local quantities entering the effective theory.

We formulate the quantum theory using the standard background-field decomposition,
\bea
\Phi(x)&=&\varphi_c(x)+\widetilde{\varphi}(x),
\label{eq:background_field_decomposition_main}
\eea
where $\varphi_c(x)$ denotes the classical background field and $\widetilde{\varphi}(x)$ represents the quantum fluctuations. This decomposition provides the usual framework for describing spontaneous symmetry breaking through the dynamics of the background field. In the present analysis, we adopt the local-potential approximation and consider a constant background,
\bea
\partial_\mu\varphi_c&=&0.
\label{eq:constant_background_main}
\eea
This assumption considerably simplifies the treatment of the effective potential while allowing the propagator of the fluctuations to retain the spatial dependence generated by the finite rotating geometry. A fully general treatment would require allowing the order parameter to vary spatially and retaining $\varphi_c=\varphi_c(r)$ in the effective action. Such an inhomogeneous background-field treatment lies beyond the present approximation and is left for future work.

For the constant background considered here, the background dependence of the fluctuation spectrum is contained in the effective mass,
\bea
m_{\mathrm{eff}}^2&=&m^2+\frac{\lambda}{2}\varphi_c^2,
\label{eq:effective_mass_main}
\eea
and the corresponding mode energy is
\bea
\omega_{l,h,c}(k_z)&=&\sqrt{k_z^2+k_{\perp,l,h}^2+m_{\mathrm{eff}}^2}.
\label{eq:mode_energy_main}
\eea

At finite temperature, the propagator is obtained from the zero-temperature expression through the imaginary-time formalism. The effect of rotation enters through the shift of the bosonic Matsubara frequencies, $\omega_n\rightarrow\omega_n-\ii l\Omega$, while the radial structure remains encoded in the Fourier-Bessel modes. The corresponding Matsubara summation and coincidence limit are derived in Appendix~\ref{app:functional-proofs}, yielding the local thermal propagator given in Eq.~\eqref{eq:app_thermal_propagator}. The thermal contribution to the self-energy follows from the corresponding tadpole, and its given by
\bea
\Sigma_T(r,\Omega,T;\varphi_c)&=&\frac{\lambda}{2}G_T(r,\Omega,T;\varphi_c).
\label{eq:thermal_self_energy_main}
\eea
Since $G_T$ depends on the radial coordinate, the thermal self-energy is likewise spatially dependent. In the Fourier-Bessel basis, $\Sigma_T$ remains diagonal in the azimuthal quantum number $l$, whereas its projection onto the radial modes is generally non-diagonal. The thermal self-energy therefore couples different radial modes within a given angular-momentum sector, a feature that must be retained when constructing the ring-resummed propagator.

The one-loop effective potential is obtained by combining the classical, vacuum, and thermal contributions derived in Appendix~\ref{app:functional-proofs}. The infrared-sensitive thermal contributions are then incorporated through the ring resummation. For the spatially dependent thermal self-energy considered here, the resummation is formulated in terms of the interpolating operator and propagator introduced in Appendix~\ref{app:ring-resummation}. The resulting local ring contribution, with the subtraction required to avoid double counting of the one-loop tadpole contribution, is given in Eq.~\eqref{eq:app_ring_local_potential}. Its matrix representation in the Fourier-Bessel basis and the corresponding spectral formulation are developed in Appendix~\ref{app:ring-resummation}.

For a constant background field, the effective action can be expressed in terms of a spatially resolved effective potential density,
\bea
V_{\mathrm{eff}}(\varphi_c,r)&=&V^{(0)}(\varphi_c)+V_{\mathrm{vac}}^{(1)}(\varphi_c,r)\nn\\
&+&V_T^{(1)}(\varphi_c,r,\Omega,T)+V_{\mathrm{ring}}(\varphi_c,r,\Omega,T),\nn\\
\label{eq:fully_resummed_local_potential-main}
\eea
where $V^{(0)}$ denotes the classical contribution, $V_{\mathrm{vac}}^{(1)}$ and $V_T^{(1)}$ are the vacuum and thermal one-loop contributions, respectively, and $V_{\mathrm{ring}}$ is the thermal ring correction. Although the background field $\varphi_c$ is taken to be constant, the effective potential density retains an explicit dependence on the radial coordinate through the spatial structure of the finite rotating system. This local quantity therefore describes how confinement and rotation affect the effective potential across the transverse plane. The effective action is obtained by integrating this density over the spatial volume, which motivates the definition of the volume averaged effective potential,
\bea
\overline{V}_{\mathrm{eff}}(\varphi_c)&=&\frac{2}{R^2}\int_0^Rdr\,r\,V_{\mathrm{eff}}(\varphi_c,r).
\label{eq:volume_averaged_effective_potential-main}
\eea
The volume average provides a single function of the constant background field and is the quantity used to characterize the global thermodynamic behavior of the finite system. Its equilibrium configurations are determined from
\bea
\frac{\partial\overline{V}_{\mathrm{eff}}}{\partial\varphi_c}&=&0.
\label{eq:global_stationarity_main}
\eea

The volume averaged effective potential $\overline{V}_{\mathrm{eff}}(\varphi_c)$ will be used to characterize the equilibrium structure of the system and to investigate the restoration of the $\mathbb{Z}_2$ symmetry as a function of temperature, rotation, and system size.

\section{Results and discussion}\label{sec:results}

The symmetry-restoration temperature is determined from the spatially averaged effective potential obtained from the fully resummed local effective potential in Eq.~\eqref{eq:fully_resummed_local_potential-main}. Following the criterion for a continuous $\mathbb{Z}_2$ symmetry-restoring transition, $T_c$ is identified from the vanishing of the curvature at the symmetric configuration, $\varphi_c=0$, together with the disappearance of the nontrivial minima associated with the broken-symmetry phase. We evaluate this condition using the high-temperature approximation discussed in Appendix~\ref{app:high_temp_expansion}. The resulting critical temperature, $T_c(\Omega,R)$, is studied as a function of the angular velocity and transverse size. In the numerical evaluation, the mode sums contain $12280$ terms, and the critical temperature is determined with a numerical tolerance of $5\times10^{-3}$.

\begin{figure}[h!]
    \centering
    \includegraphics[width=0.48\textwidth]{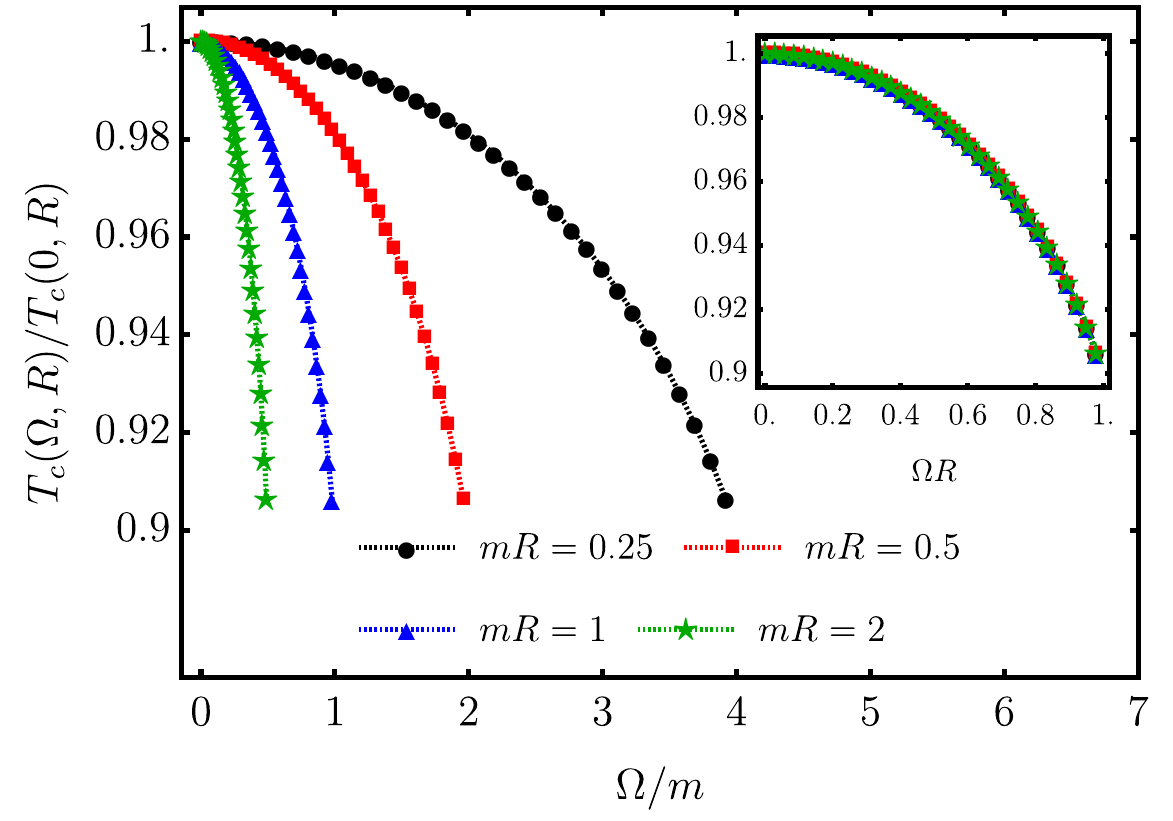}
    \caption{Normalized critical temperature for symmetry restoration, $T_c(\Omega,R)/T_c(0,R)$, as a function of $\Omega/m$ for different values of $mR$. The inset displays the same results as a function of the dimensionless boundary velocity $\Omega R$, revealing a common scaling behavior. The upper limit $\Omega R=1$ corresponds to the causality condition at the boundary~\cite{PhysRevD.108.085016}.}
    \label{fig:TcVsOmega}
\end{figure}

Figure~\ref{fig:TcVsOmega} shows the dependence of the normalized symmetry-restoration temperature on the angular velocity for several values of the dimensionless transverse size $mR$. The ratio $T_c(\Omega,R)/T_c(0,R)$ decreases as $\Omega/m$ increases, with a stronger reduction at larger $mR$ over the range of angular velocities common to the different systems. The allowed range of $\Omega$ is restricted by the causal condition at the boundary, $\Omega R<1$, which gives $\Omega<4m$, $\Omega<2m$, $\Omega<m$, and $\Omega<m/2$ for $mR=0.25$, $0.5$, $1$, and $2$, respectively. The different endpoints of the curves therefore reflect the corresponding light-cylinder bound rather than a dynamical cutoff. At fixed $\Omega/m$, the value of $\Omega R$ increases with $mR$, accounting for the different positions sampled along the rotational dependence.

The inset provides a different representation of the same numerical results, using $\Omega R$ as the horizontal variable. The curves obtained for the different transverse sizes collapse onto a common behavior. This indicates that, within the range explored, the relative rotational modification of the transition temperature is organized by the dimensionless boundary velocity. The scaling can be written as
\bea
T_c(\Omega,R)&=&F(\Omega R)T_c(0,R),
\label{scalling}
\eea
where the dependence on the transverse size remains in the nonrotating critical temperature and in the finite-volume spectrum, while the normalized rotational modification is described by the same function of $\Omega R$. This behavior is obtained from the fully resummed effective potential and therefore remains after the position-dependent thermal self-energy has been incorporated through ring resummation. It is also consistent with the scaling structure identified for rigidly rotating scalar fields in finite cylindrical systems by Ambru{\ifmmode \mbox{\c{s}}\else \c{s}\fi{}} et al.~\cite{PhysRevD.108.085016}. The present result shows that this dependence remains visible in the interacting calculation considered here.

The scaling does not imply that the finite transverse size drops out of the thermal dynamics. The discrete spectrum remains explicitly dependent on $R$ through $k_{\perp,l,h}=\gamma_{l,h}/R$, while the thermal mode energies contain the rotational shifts $\omega_{l,h,c}\pm l\Omega$. These quantities determine the radial structure of the thermal propagator and the associated position-dependent self-energy entering the resummed effective potential. The scaling in Eq.~\eqref{scalling} concerns the normalized change of $T_c$ relative to its value at zero rotation, rather than the full critical temperature itself. Consequently, systems with different transverse sizes can have different values of $T_c(0,R)$ and different underlying spectra while exhibiting the same relative rotational modification when their boundary velocities are equal.

The causal restriction and the thermal response consequently play distinct roles. The condition $\Omega R<1$ defines the kinematic domain in which the rigidly rotating configuration is considered, whereas the variation of $T_c$ within this domain results from the rotational modification of the thermal mode occupations together with the finite-volume spectrum and its contribution to the resummed effective potential. At fixed $\Omega/m$, changing $mR$ changes the boundary velocity and hence the point at which the system samples the rotational dependence described by Eq.~\eqref{scalling}.

For a phenomenological estimate, we take $m\sim100\,\mathrm{MeV}$ as a representative hadronic energy scale. Here, $m$ denotes the mass parameter of the effective $\lambda\phi^4$ theory and serves as the reference scale of the model, rather than the mass of a specific hadronic state. This gives $1/m\simeq1.97\,\mathrm{fm}$, so that $mR=0.25$, $0.5$, $1$, and $2$ correspond to $R\simeq0.49$, $0.99$, $1.97$, and $3.94\,\mathrm{fm}$, respectively. These transverse scales are comparable to characteristic dimensions of finite systems produced in heavy-ion collisions, whose size depends on the collision geometry and stage of the evolution~\cite{PhysRevC.59.3324,PhysRevC.83.044910,PhysRevC.93.024905}. Global $\Lambda$ hyperon polarization measurements in noncentral heavy-ion collisions provide evidence for substantial vorticity in the produced matter~\cite{star2017polarization}. Angular velocities inferred from such measurements are of order $\Omega\simeq7$--$10\,\mathrm{MeV}$, corresponding to $\Omega/m\simeq0.07$--$0.10$ for the scale adopted here. For the range of radii considered, this gives $\Omega R\simeq0.02$--$0.20$, placing the corresponding configurations well inside the causal domain. The parameter range explored here therefore lies away from the light-cylinder boundary while remaining within scales relevant to finite rotating systems.

The calculation thus combines the discrete finite-volume spectrum, rotational modification of the thermal modes, and the spatially dependent thermal self-energy within a resummed effective-potential framework. The resulting dependence of $T_c(\Omega,R)$ on the transverse size and angular velocity is accompanied by the scaling behavior in Eq.~\eqref{scalling}, which survives the inclusion of the ring contributions and organizes the normalized rotational response in terms of the boundary velocity.

\section{Summary and conclusions}\label{sec:conclusions}

We have analyzed $\mathbb{Z}_2$ symmetry restoration in a rotating $\lambda\phi^4$ field confined to a finite cylindrical geometry, incorporating finite transverse confinement, thermal fluctuations, and rotation within a resummed effective potential. The finite boundary conditions generate a discrete Fourier-Bessel spectrum, while rigid rotation modifies the thermal occupation of the angular-momentum modes. The resulting thermal propagator depends explicitly on the radial coordinate, giving rise to a position-dependent thermal self-energy that enters the effective potential through ring resummation. This formulation retains the spatial structure induced by the finite geometry up to the determination of the spatially averaged effective potential and the corresponding symmetry-restoration temperature.

The critical temperature $T_c(\Omega,R)$ depends on both the angular velocity and the transverse size. The finite radius determines the transverse eigenvalues and their spacing, which modify the mode energies entering the thermal propagator and, consequently, the spatial profile of the thermal self-energy. These contributions are incorporated into the resummed effective potential and modify the condition for the disappearance of the broken-symmetry minimum. The dependence on $R$ therefore originates from the finite-volume spectral structure of the rotating thermal system and is present already at the level of the fluctuation modes.

The normalized critical temperature exhibits a scaling behavior described by Eq.~\eqref{scalling}. Two systems with different transverse sizes yield the same relative modification of the critical temperature when their boundary velocities $\Omega R$ are equal, despite having different transverse spectra. The scaling therefore emerges after the discrete mode structure and the position-dependent thermal self-energy have been incorporated into the effective potential. Its persistence in the presence of ring resummation connects the present interacting calculation with the $\Omega R$ scaling found previously for rigidly rotating scalar systems~\cite{PhysRevD.108.085016}.

The causal condition $\Omega R<1$ restricts the admissible rotational states but does not determine the thermal response within this domain. The latter follows from the occupation of the discrete modes and their contribution to the position-dependent self-energy. The light-cylinder condition consequently defines the kinematic domain of the rotating configuration, while the finite-volume spectrum determines the thermal response within that domain.

The resulting framework provides a field-theoretical description in which finite transverse confinement, rotation, and thermal resummation are treated simultaneously while retaining the spatial structure of the fluctuations. In particular, the position dependence of the thermal self-energy allows the radial inhomogeneity generated by the finite geometry to enter directly into the effective potential before the final spatial averaging. This makes it possible to determine how changes in the transverse size modify the symmetry-restoration temperature while preserving the discrete spectral information of the confined system.

The phenomenological estimates considered above place the parameters explored here within a regime well inside the causal domain, with transverse sizes and angular velocities compatible with scales characteristic of finite rotating systems. The resulting $T_c(\Omega,R)$ dependence and its scaling with $\Omega R$ provide a basis for investigating how these features change when additional dynamical degrees of freedom, alternative boundary conditions, or corrections beyond the high-temperature approximation are included.

\section*{Acknowledgments} 
LAC-S dedicates this work to the memory of his grandmother, Betty, whom he lovingly called {\it mami}. He carries her memory with him every day and promised that her name would accompany him throughout his scientific career, appearing in every work he writes. JDC-Y would like to thank the Isaac Newton Institute for Mathematical Sciences (INI), Cambridge, for support and hospitality during the programme \textit{Quantum field theory with boundaries, impurities, and defects (BID2025)} where work on this paper was undertaken. JDC-Y would also like to thank the Vicerrectoría de Investigaciones of Universidad del Valle, Colombia, for supporting his medium-term research stay at the INI, which allowed part of this project to be carried out. The authors would also like to thank Maxim Chernodub, Institut Denis Poisson, Université de Tours, France, for valuable comments on the scaling behavior identified in the results.

\section*{Declaration on the Use of Artificial Intelligence}

During the preparation of this work, artificial intelligence tools, including ChatGPT and Gemini, were used as aids for writing, proofreading, and scientific cross-checking. Their use included language editing, identification of typographical errors, and consistency checks of equations, signs, notation, and mathematical expressions. These tools were not used to independently generate the scientific results or conclusions of this work. All scientific content, calculations, and interpretations were reviewed and verified by the authors, who take full responsibility for the content of the manuscript.

\appendix
\section{Geometry of the Rotating Cylindrical Frame}\label{sec:rotating_geometry}

We formulate the problem in cylindrical coordinates $(t,r,\phi,z)$ associated with a frame rotating uniformly about the $z$ axis. We take the inertial Minkowski coordinates to be $(t',r',\phi',z')$, with
\bea
(t',r',\phi',z')\to(t,r,\phi+\Omega t,z),
\label{eq:rotating_coordinate_transformation}
\eea
where $\Omega$ is the constant angular velocity measured with respect to the inertial frame. Starting from the Minkowski line element in cylindrical coordinates,
\bea
ds^2=dt'^2-dr'^2-r'^2d\phi'^2-dz'^2,
\label{eq:minkowski_cylindrical_metric}
\eea
the coordinate transformation in Eq.~\eqref{eq:rotating_coordinate_transformation} gives
\bea
ds^2=\left(1-\Omega^2r^2\right)dt^2-2\Omega r^2dt\,d\phi-dr^2-r^2d\phi^2-dz^2.\nn\\
\label{eq:rotating_line_element}
\eea
The corresponding covariant metric is
\bea
g_{\mu\nu}=
\begin{pmatrix}
1-\Omega^2r^2&0&-\Omega r^2&0\\
0&-1&0&0\\
-\Omega r^2&0&-r^2&0\\
0&0&0&-1
\end{pmatrix},
\label{eq:rotating_covariant_metric}
\eea
where the coordinate ordering is $(t,r,\phi,z)$. Its determinant and associated volume element are
\bea
g=\det(g_{\mu\nu})=-r^2,
\label{eq:rotating_metric_determinant}
\eea

The inverse metric follows from $g^{\mu\alpha}g_{\alpha\nu}=\delta^\mu_{\ \nu}$ and is given by
\bea
g^{\mu\nu}=
\begin{pmatrix}
1&0&-\Omega&0\\
0&-1&0&0\\
-\Omega&0&\Omega^2-\frac{1}{r^2}&0\\
0&0&0&-1
\end{pmatrix}.
\label{eq:rotating_contravariant_metric}
\eea
The off-diagonal component $g_{0\phi}=-\Omega r^2$ encodes the mixing between time translations and azimuthal rotations induced by the choice of rotating coordinates. 

Although the metric in Eq.~\eqref{eq:rotating_covariant_metric} is coordinate dependent and contains an off-diagonal temporal-azimuthal component, the underlying spacetime remains flat. This follows from the fact that the rotating coordinates are related to inertial Minkowski coordinates by a coordinate transformation. The Riemann curvature tensor, defined in terms of the Levi-Civita connection as
\bea
R^\mu_{\ \nu\alpha\beta}=\partial_\alpha\Gamma^\mu_{\beta\nu}-\partial_\beta\Gamma^\mu_{\alpha\nu}+\Gamma^\mu_{\alpha\lambda}\Gamma^\lambda_{\beta\nu}-\Gamma^\mu_{\beta\lambda}\Gamma^\lambda_{\alpha\nu},
\label{eq:rotating_riemann_tensor}
\eea
vanishes for the metric in Eq.~\eqref{eq:rotating_covariant_metric}. Its contractions, namely the Ricci tensor,
\bea
R_{\mu\nu}=R^\alpha_{\ \mu\alpha\nu},
\label{eq:rotating_ricci_tensor}
\eea
and the Ricci scalar,
\bea
R=g^{\mu\nu}R_{\mu\nu},
\label{eq:rotating_ricci_scalar}
\eea
also vanish,
\bea
R^\mu_{\ \nu\alpha\beta}=0,\qquad R_{\mu\nu}=0,\qquad R=0.
\label{eq:rotating_flat_spacetime}
\eea
The nontrivial form of the metric therefore reflects the non-inertial character of the rotating coordinates rather than a nonzero spacetime curvature.

The rotating coordinates are well defined only inside the light cylinder. To see this directly, consider a trajectory with fixed $(r,\phi,z)$. Along such a worldline,
\bea
ds^2=\left(1-\Omega^2r^2\right)dt^2.
\label{eq:rotating_stationary_worldline}
\eea
The coefficient of $dt^2$ changes sign at
\bea
r_{\mathrm{LC}}=\frac{1}{\Omega},
\label{eq:rotating_light_cylinder}
\eea
which corresponds to the radius at which the tangential velocity of a co-rotating observer, $v=\Omega r$, reaches unity. We therefore restrict the system to a finite cylindrical region of radius $R$ satisfying
\bea
\Omega R<1.
\label{eq:rotating_causality_condition}
\eea
This condition places the entire spatial domain inside the light cylinder and will be assumed in the construction of the scalar spectrum.

\section{Scalar Propagator in the Rotating Cylindrical System}
\label{app:scalar_propagator}

We consider a real scalar field of mass $m$ in the uniformly rotating cylindrical system introduced in Appendix~\ref{sec:rotating_geometry}. The field is defined inside a cylinder of radius $R$, with the angular velocity restricted by the causality condition $\Omega R<1$. The scalar field satisfies the covariant Klein--Gordon equation
\bea
\left(\Box_g+m^2\right)\Phi(x)=0,
\label{eq:app_covariant_KG}
\eea
where
\bea
\Box_g=\frac{1}{\sqrt{-g}}\partial_\mu\left(\sqrt{-g}g^{\mu\nu}\partial_\nu\right).
\label{eq:app_dalembertian}
\eea
For the rotating cylindrical metric of Appendix~\ref{sec:rotating_geometry}, one has $\sqrt{-g}=r$. The Klein--Gordon equation can therefore be written explicitly as
\bea
\bigg[\left(\ii\frac{\partial}{\partial t}+\Omega\hat L_z\right)^2+\nabla^2-m^2\bigg]\Phi(x)=0,
\label{eq:app_rotating_KG}
\eea
where $\hat L_z=-\ii\partial_\phi$ and
\bea
\nabla^2=\frac{\partial^2}{\partial r^2}+\frac{1}{r}\frac{\partial}{\partial r}+\frac{1}{r^2}\frac{\partial^2}{\partial\phi^2}+\frac{\partial^2}{\partial z^2}.
\label{eq:app_cylindrical_laplacian}
\eea
We impose a Dirichlet boundary condition at the cylindrical surface,
\bea
\Phi(t,R,\phi,z)=0.
\label{eq:app_dirichlet_boundary}
\eea
This condition turns the radial part of the Klein--Gordon equation into a finite-interval boundary-value problem and leads to a discrete transverse spectrum.

To construct the corresponding normal modes, we use the separation ansatz
\bea
\Phi_{E,k_z,l}(x)=e^{-\ii Et}e^{\ii k_z z}e^{\ii l\phi}f_l(r),
\label{eq:app_separation_ansatz}
\eea
where $E$ is the frequency conjugate to the rotating time coordinate, $k_z$ is the momentum along the rotation axis, and $l\in\mathbb{Z}$ is the azimuthal quantum number. Since
\bea
\hat L_z e^{\ii l\phi}=l e^{\ii l\phi},
\label{eq:app_angular_eigenvalue}
\eea
substitution of Eq.~\eqref{eq:app_separation_ansatz} into Eq.~\eqref{eq:app_rotating_KG} gives
\bea
\left[\frac{d^2}{dr^2}+\frac{1}{r}\frac{d}{dr}+(E+\Omega l)^2-k_z^2-m^2-\frac{l^2}{r^2}\right]f_l(r)=0.
\label{eq:app_radial_equation}
\eea
It is convenient to introduce the transverse momentum via
\bea
k_\perp^2=(E+\Omega l)^2-k_z^2-m^2,
\label{eq:app_kperp_definition}
\eea
so that Eq.~\eqref{eq:app_radial_equation} becomes
\bea
\left[\frac{d^2}{dr^2}+\frac{1}{r}\frac{d}{dr}+\left(k_\perp^2-\frac{l^2}{r^2}\right)\right]f_l(r)=0.
\label{eq:app_bessel_equation}
\eea
This is Bessel's equation of order $|l|$. The two linearly independent solutions are the Bessel functions of the first and second kinds. Regularity at the symmetry axis excludes the second kind, since $Y_l(k_\perp r)$ is singular at $r=0$. The admissible radial solution is therefore
\bea
f_l(r)\propto J_l(k_\perp r).
\label{eq:app_regular_radial_solution}
\eea
The Dirichlet condition in Eq.~\eqref{eq:app_dirichlet_boundary} then requires
\bea
J_l(k_\perp R)=0.
\label{eq:app_bessel_boundary}
\eea
Let $\gamma_{l,h}$ denote the $h$th positive zero of $J_l$,
\bea
J_l(\gamma_{l,h})=0,\qquad h=1,2,\ldots.
\label{eq:app_bessel_zeros}
\eea
The allowed transverse momenta are consequently quantized according to
\bea
k_{\perp,l,h}=\frac{\gamma_{l,h}}{R}.
\label{eq:app_discrete_transverse_momentum}
\eea
The complete set of separated modes can thus be written as
\bea
\Phi_{E,k_z,l,h}(x)=\mathcal{N}_{l,h}e^{-\ii Et}e^{\ii k_z z}e^{\ii l\phi}J_l(k_{\perp,l,h}r),
\label{eq:app_scalar_modes}
\eea
where $\mathcal{N}_{l,h}$ denotes the normalization factor associated with the radial, angular, and longitudinal mode functions.

Substitution of the quantized transverse momentum into Eq.~\eqref{eq:app_kperp_definition} gives the mode dispersion relation
\bea
(E+\Omega l)^2=k_z^2+k_{\perp,l,h}^2+m^2.
\label{eq:app_mode_dispersion}
\eea
It is useful to define
\bea
\omega_{l,h}(k_z)=\sqrt{k_z^2+k_{\perp,l,h}^2+m^2},
\label{eq:app_mode_energy}
\eea
in terms of which the two solutions for the frequency are
\bea
E=-\Omega l\pm\omega_{l,h}(k_z).
\label{eq:app_energy_poles}
\eea
The appearance of $E+\Omega l$ is the characteristic modification introduced by the rotating coordinates and will determine the rotational dependence of the propagator.

The normalization of the radial modes follows from the orthogonality relation for Bessel functions evaluated at their zeros,
\bea
\int_0^R r\,dr\,J_l(k_{\perp,l,h}r)J_l(k_{\perp,l,h'}r)=\frac{R^2}{2}J_{l+1}^2(\gamma_{l,h})\delta_{hh'}.\nn\\
\label{eq:app_bessel_orthogonality}
\eea
We therefore introduce the normalized radial functions
\bea
U_{l,h}(r)=\frac{\sqrt{2}}{R J_{l+1}(\gamma_{l,h})}J_l(k_{\perp,l,h}r),
\label{eq:app_normalized_radial_modes}
\eea
which satisfy
\bea
\int_0^R r\,dr\,U_{l,h}(r)U_{l,h'}(r)=\delta_{hh'}.
\label{eq:app_radial_normalization}
\eea
The angular dependence is normalized with
\bea
\int_0^{2\pi}d\phi\,\frac{e^{\ii l\phi}}{\sqrt{2\pi}}\frac{e^{-\ii l'\phi}}{\sqrt{2\pi}}=\delta_{ll'},
\label{eq:app_angular_normalization}
\eea
and the longitudinal plane waves satisfy
\bea
\int_{-\infty}^{\infty}dz\,\frac{e^{\ii k_z z}}{\sqrt{2\pi}}\frac{e^{-\ii k_z'z}}{\sqrt{2\pi}}=\delta(k_z-k_z').
\label{eq:app_longitudinal_normalization}
\eea
The normalized spatial eigenfunctions can consequently be written as
\bea
\Psi_{l,h,k_z}(\mathbf{x})=U_{l,h}(r)\frac{e^{\ii l\phi}}{\sqrt{2\pi}}\frac{e^{\ii k_z z}}{\sqrt{2\pi}},
\label{eq:app_spatial_modes}
\eea
where $\mathbf{x}=(r,\phi,z)$. Their completeness relation is
\bea
&&\frac{1}{r}\delta(r-r')\delta(\phi-\phi')\delta(z-z')\nn\\
&=&\sum_{l=-\infty}^{\infty}\sum_{h=1}^{\infty}\int\frac{dk_z}{2\pi}\Psi_{l,h,k_z}(\mathbf{x})\Psi_{l,h,k_z}^*(\mathbf{x}').
\label{eq:app_spatial_completeness}
\eea
The factor $1/r$ follows from the cylindrical spatial measure $r\,dr\,d\phi\,dz$. In terms of the original Bessel functions, the radial part of the completeness relation is
\bea
\frac{1}{r}\delta(r-r')=\frac{2}{R^2}\sum_{h=1}^{\infty}\frac{J_l(k_{\perp,l,h}r)J_l(k_{\perp,l,h}r')}{J_{l+1}^2(\gamma_{l,h})}.
\label{eq:app_bessel_completeness}
\eea
This relation will determine the normalization factor appearing in the spectral representation of the propagator.

We now construct the Feynman propagator as the Green function associated with the Klein--Gordon operator. Our convention is
\bea
\left(\Box_g+m^2\right)G_\text{F}(x,x')=-\frac{1}{\sqrt{-g}}\delta^{(4)}(x-x').
\label{eq:app_propagator_definition}
\eea

Since $\sqrt{-g}=r$, the source on the right-hand side is explicitly
\bea
&&\frac{1}{\sqrt{-g}}\delta^{(4)}(x-x')\nn\\
&=&\frac{1}{r}\delta(t-t')\delta(r-r')\delta(\phi-\phi')\delta(z-z').
\label{eq:app_covariant_delta}
\eea
To invert the Klein--Gordon operator in the mode basis, it is convenient to define
\bea
\hat{\mathcal{D}}\equiv-\left(\Box_g+m^2\right)=\left(\ii\frac{\partial}{\partial t}+\Omega\hat L_z\right)^2+\nabla^2-m^2.\nn\\
\label{eq:app_spectral_operator}
\eea

The modes introduced above are eigenfunctions of this operator,
\bea
\hat{\mathcal{D}}\Phi_{E,k_z,l,h}(x)=\lambda_{E,k_z,l,h}\Phi_{E,k_z,l,h}(x),
\label{eq:app_operator_eigenvalue}
\eea
with eigenvalues
\bea
\lambda_{E,k_z,l,h}=(E+\Omega l)^2-k_z^2-k_{\perp,l,h}^2-m^2.
\label{eq:app_spectral_eigenvalue}
\eea
By Eq.~\eqref{eq:app_mode_dispersion}, the eigenvalue vanishes when the mode is on shell,
\bea
\lambda_{E,k_z,l,h}=0\quad\Longleftrightarrow\quad(E+\Omega l)^2=\omega_{l,h}^2(k_z).
\label{eq:app_on_shell_condition}
\eea

The Green function is obtained by expanding in the complete set of modes and replacing each eigenvalue by its inverse. Since the inverse is singular at the eigenvalues satisfying Eq.~\eqref{eq:app_on_shell_condition}, the Feynman prescription is implemented as
\bea
\lambda_{E,k_z,l,h}\rightarrow\lambda_{E,k_z,l,h}+\ii\epsilon,\qquad \epsilon\rightarrow0^+.
\label{eq:app_feynman_prescription}
\eea
Using the normalized modes in Eq.~\eqref{eq:app_spatial_modes}, the corresponding radial normalization in Eq.~\eqref{eq:app_normalized_radial_modes}, and the Fourier representation of the continuous variables, the propagator takes the form
\begin{widetext}
  \bea
G_\text{F}(x,x')=\frac{1}{\pi R^2}\sum_{l=-\infty}^{\infty}\sum_{h=1}^{\infty}\frac{1}{J_{l+1}^2(\gamma_{l,h})}\int\frac{dE\,dk_z}{(2\pi)^2}\frac{e^{-\ii E(t-t')}e^{\ii k_z(z-z')}e^{\ii l(\phi-\phi')}J_l(k_{\perp,l,h}r)J_l(k_{\perp,l,h}r')}{(E+\Omega l)^2-k_z^2-k_{\perp,l,h}^2-m^2+\ii\epsilon}.
\label{eq:app_scalar_propagator}
\eea  
\end{widetext}

The prefactor in Eq.~\eqref{eq:app_scalar_propagator} is fixed by the normalization of the radial and angular eigenfunctions. In particular, the factor $J_{l+1}^{-2}(\gamma_{l,h})$ originates from the radial orthogonality relation in Eq.~\eqref{eq:app_bessel_orthogonality}.

The Dirichlet boundary condition is inherited by the propagator in each of its spatial arguments. Indeed, since
\bea
J_l(k_{\perp,l,h}R)=J_l(\gamma_{l,h})=0,
\eea
the radial factors in Eq.~\eqref{eq:app_scalar_propagator} imply
\bea
G_\text{F}(x,x')\big|_{r=R}=0,
\qquad
G_\text{F}(x,x')\big|_{r'=R}=0.
\label{eq:app_propagator_boundary}
\eea
Thus, the propagator vanishes whenever either of its spatial arguments lies on the cylindrical boundary.

The behavior at the symmetry axis is determined by the regular radial solution. For $r\rightarrow0$,
\bea
J_l(k_{\perp,l,h}r)\sim r^{|l|},
\label{eq:app_bessel_origin}
\eea
whereas the second independent solution $Y_l(k_{\perp,l,h}r)$ is singular at $r=0$ and is excluded from the mode basis. The propagator therefore inherits the regularity of the scalar modes at the axis.

As a consistency check, consider the nonrotating limit $\Omega\rightarrow0$. Equation~\eqref{eq:app_scalar_propagator} becomes
\begin{widetext}
   \bea
G_\text{F}(x,x')\big|_{\Omega=0}=\frac{1}{\pi R^2}\sum_{l=-\infty}^{\infty}\sum_{h=1}^{\infty}\frac{1}{J_{l+1}^2(\gamma_{l,h})}\int\frac{dE\,dk_z}{(2\pi)^2}\frac{e^{-\ii E(t-t')}e^{\ii k_z(z-z')}e^{\ii l(\phi-\phi')}J_l(k_{\perp,l,h}r)J_l(k_{\perp,l,h}r')}{E^2-k_z^2-k_{\perp,l,h}^2-m^2+\ii\epsilon}.
\label{eq:app_nonrotating_propagator}
\eea 
\end{widetext}

At finite $R$, the transverse momentum remains discrete as a consequence of the Dirichlet boundary condition. When $R\rightarrow\infty$, the allowed values of $k_{\perp,l,h}$ become dense and the weighted sum over the radial quantum number approaches the continuum transverse-momentum measure,
\bea
\frac{1}{\pi R^2}\sum_{h=1}^{\infty}\frac{F(k_{\perp,l,h})}{J_{l+1}^2(\gamma_{l,h})}\longrightarrow\int_0^\infty\frac{k_\perp dk_\perp}{2\pi}F(k_\perp),
\label{eq:app_continuum_limit}
\eea
for functions $F$ for which the continuum limit exists. The sum over $l$ can then be combined with the Bessel functions via the cylindrical completeness relation, yielding the translationally invariant Minkowski propagator,
\bea
G_\text{F}(x-x')=\int\frac{d^4p}{(2\pi)^4}\frac{e^{-\ii p\cdot(x-x')}}{p^2-m^2+\ii\epsilon}.
\label{eq:app_minkowski_limit}
\eea
The $R\rightarrow\infty$ limit therefore recovers the standard continuum normalization and provides a consistency check of the finite-cylinder spectral representation.

\section{Functional Derivation of the One-Loop Effective Action}\label{app:functional-proofs}

For completeness, we review the functional construction of the effective action and its derivative expansion. The local effective potential is identified with the zero-derivative contribution to the effective action, while the derivative terms account for the momentum dependence of the background field.

We consider the generating functional
\bea
Z[J]&=&\int\mathcal{D}\varphi\,\exp\left\{\ii S[\varphi]+\ii\int d^4x\,J(x)\varphi(x)\right\},\nn\\
\eea
and define the connected generating functional $W[J]$ via
\bea
Z[J]&=&\exp\left\{\ii W[J]\right\}.
\eea
The classical field is obtained from
\bea
\varphi_c(x)&=&\frac{\delta W[J]}{\delta J(x)},
\eea
while the effective action is defined by the Legendre transformation
\bea
\Gamma[\varphi_c]&=&W[J]-\int d^4x\,J(x)\varphi_c(x).
\eea
Consequently,
\bea
\frac{\delta\Gamma[\varphi_c]}{\delta\varphi_c(x)}&=&-J(x).
\eea

The semiclassical expansion is obtained by decomposing the field into a background configuration and a fluctuation,
\bea
\varphi(x)&=&\varphi_c(x)+\eta(x).
\eea
The background field satisfies the stationary condition in the presence of the external source,
\bea
\left.\frac{\delta S[\varphi]}{\delta\varphi(x)}\right|_{\varphi=\varphi_c}+J(x)&=&0.
\label{eq:saddle_point_condition}
\eea
Expanding the action around $\varphi_c$ gives
\bea
S[\varphi_c+\eta]&=&S[\varphi_c]+\int d^4x\,\left.\frac{\delta S[\varphi]}{\delta\varphi(x)}\right|_{\varphi_c}\eta(x)\nn\\
&&+\frac{1}{2}\int d^4x\,d^4y\,\eta(x)K(x,y;\varphi_c)\eta(y)\nn\\
&&+\mathcal{O}(\eta^3),
\eea
where the quadratic fluctuation operator is defined by
\bea
K(x,y;\varphi_c)&=&\left.\frac{\delta^2S[\varphi]}{\delta\varphi(x)\delta\varphi(y)}\right|_{\varphi=\varphi_c}.
\label{eq:quadratic_fluctuation_operator}
\eea
The source term is expanded as
\bea
\int d^4x\,J(x)\varphi(x)&=&\int d^4x\,J(x)\varphi_c(x)\nn\\
&&+\int d^4x\,J(x)\eta(x).
\eea
Using Eq.~\eqref{eq:saddle_point_condition}, the terms linear in $\eta$ cancel, and the generating functional becomes
\bea
Z[J]&\simeq&\exp\left\{\ii S[\varphi_c]+\ii\int d^4x\,J(x)\varphi_c(x)\right\}\nn\\
&&\quad\times\int\mathcal{D}\eta\,\exp\bigg\{\frac{\ii}{2}\int d^4x\,d^4y\,\eta(x)\nn\\
&&\qquad\times K(x,y;\varphi_c)\eta(y)\bigg\},
\eea
where terms cubic and higher order in the fluctuation field generate contributions beyond the Gaussian approximation. At one-loop order, the functional integral is therefore determined by the quadratic fluctuation operator. Up to a field-independent normalization, one has
\bea
&&\int\mathcal{D}\eta\,\exp\left\{\frac{\ii}{2}\int d^4x\,d^4y\,\eta(x)K(x,y;\varphi_c)\eta(y)\right\}\nn\\
&\propto&\left[\det K[\varphi_c]\right]^{-1/2}=\exp\left\{-\frac{1}{2}\operatorname{Tr}\ln K[\varphi_c]\right\}.\nn\\
\eea
Thus,
\bea
Z[J]&\simeq&\exp\bigg\{\ii S[\varphi_c]+\ii\int d^4x\,J(x)\varphi_c(x)\nn\\
&&-\frac{1}{2}\operatorname{Tr}\ln K[\varphi_c]\bigg\},
\eea
and, using $Z[J]=\exp\{\ii W[J]\}$,
\bea
W[J]&=&S[\varphi_c]+\int d^4x\,J(x)\varphi_c(x)\nn\\
&&+\frac{\ii}{2}\operatorname{Tr}\ln K[\varphi_c].
\eea
Substitution into the Legendre transformation yields
\bea
\Gamma[\varphi_c]&=&S[\varphi_c]+\frac{\ii}{2}\operatorname{Tr}\ln K[\varphi_c]+\mathcal{O}(\text{two-loop}),\nn\\
\label{eq:one_loop_effective_action}
\eea
so that the one-loop contribution is
\bea
\Gamma^{(1)}[\varphi_c]&=&\frac{\ii}{2}\operatorname{Tr}\ln K[\varphi_c].
\label{eq:one_loop_determinant}
\eea

For the real $\lambda\phi^4$ theory considered in this work,
\bea
S[\varphi]&=&\int d^4x\,\left[\frac{1}{2}\partial_\mu\varphi\,\partial^\mu\varphi-\frac{1}{2}m^2\varphi^2-\frac{\lambda}{4!}\varphi^4\right],\nn\\
\eea
the quadratic fluctuation operator evaluated at the background field is
\bea
K(x,y;\varphi_c)&=&\left[-\Box-m^2-\frac{\lambda}{2}\varphi_c^2(x)\right]\delta^{(4)}(x-y).\nn\\
\label{eq:quadratic_operator}
\eea
Consequently,
\bea
\Gamma^{(1)}[\varphi_c]&=&\frac{\ii}{2}\operatorname{Tr}\ln\left[-\Box-m^2-\frac{\lambda}{2}\varphi_c^2\right].
\label{eq:one_loop_scalar}
\eea

The relation between the functional determinant and the one-loop tadpole follows by varying the determinant with respect to the background field. Using the identity $\delta\operatorname{Tr}\ln K=\operatorname{Tr}(K^{-1}\delta K)$, we obtain
\bea
\delta\Gamma^{(1)}[\varphi_c]&=&\frac{\ii}{2}\operatorname{Tr}\left[K^{-1}[\varphi_c]\delta K[\varphi_c]\right].
\label{eq:trace_log_variation}
\eea
In coordinate space,
\bea
\delta\Gamma^{(1)}[\varphi_c]&=&\frac{\ii}{2}\int d^4u\,d^4v\,K^{-1}(u,v;\varphi_c)\delta K(v,u;\varphi_c),\nn\\
\eea
and therefore
\bea
\frac{\delta\Gamma^{(1)}[\varphi_c]}{\delta\varphi_c(x)}&=&\frac{\ii}{2}\int d^4u\,d^4v\,K^{-1}(u,v;\varphi_c)\frac{\delta K(v,u;\varphi_c)}{\delta\varphi_c(x)}.\nn\\
\eea
From Eq.~\eqref{eq:quadratic_operator},
\bea
K(y,z;\varphi_c)&=&-\left[\Box_y+m^2+\frac{\lambda}{2}\varphi_c^2(y)\right]\delta^{(4)}(y-z),\nn\\
\eea
and its variation is
\bea
\frac{\delta K(v,u;\varphi_c)}{\delta\varphi_c(x)}&=&-\lambda\varphi_c(v)\delta^{(4)}(v-u)\delta^{(4)}(v-x).\nn\\
\label{eq:quadratic_operator_variation}
\eea
The inverse quadratic operator is defined via
\bea
\int d^4z\,K(x,z;\varphi_c)K^{-1}(z,y;\varphi_c)&=&\delta^{(4)}(x-y).
\label{eq:quadratic_operator_inverse}
\eea
The background-dependent Feynman propagator satisfies
\bea
\left[\Box_x+m^2+\frac{\lambda}{2}\varphi_c^2(x)\right]G_{\mathrm{F}}(x,y;\varphi_c)&=&-\delta^{(4)}(x-y).\nn\\
\label{eq:background_propagator_equation}
\eea
Since the quadratic operator in Eq.~\eqref{eq:quadratic_operator} is the negative of the differential operator appearing in Eq.~\eqref{eq:background_propagator_equation}, one has
\bea
\int d^4z\,K(x,z;\varphi_c)G_{\mathrm{F}}(z,y;\varphi_c)&=&\delta^{(4)}(x-y),
\eea
and hence
\bea
K^{-1}(x,y;\varphi_c)&=&G_{\mathrm{F}}(x,y;\varphi_c).
\label{eq:K_inverse_propagator}
\eea
Substitution into the functional derivative of the determinant gives
\bea
\frac{\delta\Gamma^{(1)}[\varphi_c]}{\delta\varphi_c(x)}&=&-\frac{\ii\lambda}{2}\varphi_c(x)G_{\mathrm{F}}(x,x;\varphi_c).
\label{eq:effective_action_tadpole_relation}
\eea
This relation provides the one-loop connection between the effective action and the tadpole evaluated in the background field.

The effective action is generally a nonlocal functional of the background field. Its local structure can be organized via a derivative expansion in powers of derivatives of $\varphi_c$. The corresponding expansion separates the zero-derivative contribution from terms containing gradients of the background field. In the finite rotating system, the coefficient functions of this expansion may depend explicitly on the spacetime coordinates because the cylindrical boundary and the rotating geometry break translational invariance.

The functional Taylor expansion in terms of one-particle-irreducible vertex functions is
\bea
\Gamma[\varphi_c]&=&\Gamma[0]+\sum_{n=1}^{\infty}\frac{1}{n!}\int\prod_{i=1}^n d^4x_i\,\Gamma^{(n)}(x_1,\ldots,x_n)\nn\\
&&\qquad\qquad\times\varphi_c(x_1)\cdots\varphi_c(x_n).
\label{eq:app_vertex_expansion}
\eea
Introducing a reference point $x$ and relative coordinates $y_i=x_i-x$, the background field can be expanded as
\bea
\varphi_c(x+y_i)&=&\varphi_c(x)+y_i^\mu\partial_\mu\varphi_c(x)\nn\\
&&+\frac{1}{2}y_i^\mu y_i^\nu\partial_\mu\partial_\nu\varphi_c(x)+\mathcal{O}(\partial^3).
\label{eq:app_background_taylor}
\eea
The terms containing no derivatives of the background field define the local zero-derivative contribution to the effective action,
\bea
\Gamma^{(0)}[\varphi_c]&\equiv&-\int d^4x\,V_{\mathrm{eff}}(\varphi_c,x).
\label{eq:app_zero_derivative}
\eea
The coordinate dependence of $V_{\mathrm{eff}}(\varphi_c,x)$ originates from the explicit spatial dependence of the coefficient functions generated by the finite geometry and does not require derivatives of the background field.

At first order in derivatives, the local bulk effective action may contain a term of the form
\bea
\Gamma_{\partial}[\varphi_c]&=&\int d^4x\,V^\mu(\varphi_c,x)\partial_\mu\varphi_c(x).
\label{eq:app_linear_derivative_general}
\eea
Define
\bea
F^\mu(\varphi_c,x)&=&\int^{\varphi_c}d\chi\,V^\mu(\chi,x),
\label{eq:app_F_definition}
\eea
so that
\bea
\frac{\partial F^\mu(\varphi_c,x)}{\partial\varphi_c}&=&V^\mu(\varphi_c,x).
\eea
The chain rule then gives
\bea
\partial_\mu F^\mu(\varphi_c,x)&=&V^\mu(\varphi_c,x)\partial_\mu\varphi_c+\left.\partial_\mu F^\mu(\varphi_c,x)\right|_{\varphi_c},\nn\\
\label{eq:app_F_chain_rule}
\eea
and therefore
\bea
V^\mu(\varphi_c,x)\partial_\mu\varphi_c&=&\partial_\mu F^\mu(\varphi_c,x)-\left.\partial_\mu F^\mu(\varphi_c,x)\right|_{\varphi_c}.\nn\\
\label{eq:app_linear_term_decomposition}
\eea
The first term on the right-hand side contributes via the surface of the integration domain, whereas the second is a local term without derivatives of the background field. Thus, up to surface contributions, a term linear in derivatives of a single scalar background field does not constitute an independent bulk structure.

At second order in derivatives, the independent bulk contribution can be expressed in terms of a kinetic functional,
\bea
\Gamma_{\mathrm{kin}}[\varphi_c]&=&\frac{1}{2}\int d^4x\,Z^{\mu\nu}(\varphi_c,x)\partial_\mu\varphi_c\partial_\nu\varphi_c+\mathcal{O}(\partial^3).\nn\\
\label{eq:app_kinetic_functional}
\eea
The tensor $Z^{\mu\nu}$ contains information about the momentum dependence of the 1PI two-point function. In the rotating finite system, the rotation axis and the cylindrical boundary distinguish different spacetime directions, so $Z^{\mu\nu}$ need not be proportional to the Minkowski metric. Terms involving second derivatives of the background can likewise be reduced to the kinetic sector and the previously discussed linear-derivative sector by integration by parts. For a general coefficient $\widetilde{Z}^{\mu\nu}(\varphi_c,x)$,
\bea
&&\int d^4x\,\widetilde{Z}^{\mu\nu}(\varphi_c,x)\varphi_c\partial_\mu\partial_\nu\varphi_c\nn\\
&=&-\int d^4x\,\widetilde{Z}^{\mu\nu}(\varphi_c,x)\partial_\mu\varphi_c\partial_\nu\varphi_c\nn\\
&&-\int d^4x\,\left.\partial_\mu\widetilde{Z}^{\mu\nu}(\varphi_c,x)\right|_{\varphi_c}\varphi_c\partial_\nu\varphi_c\nn\\
&&-\int d^4x\,\frac{\partial\widetilde{Z}^{\mu\nu}}{\partial\varphi_c}\varphi_c\partial_\mu\varphi_c\partial_\nu\varphi_c,
\label{eq:app_second_derivative_reduction}
\eea
where surface contributions have been omitted. The second term has the linear-derivative form discussed above, while the first and third terms contribute to the derivative sector.

Consequently, the bulk effective action via second order in derivatives can be written as
\bea
\Gamma[\varphi_c]&=&-\int d^4x\,V_{\mathrm{eff}}(\varphi_c,x)+\Gamma_{\mathrm{kin}}[\varphi_c]+\mathcal{O}(\partial^3).\nn\\
\label{eq:app_inhomogeneous_effective_action}
\eea
Here, $V_{\mathrm{eff}}(\varphi_c,x)$ denotes the local zero-derivative contribution, while $\Gamma_{\mathrm{kin}}$ contains the independent bulk terms involving derivatives of the background field.

The effective potential is obtained by evaluating the effective action for a constant background field,
\bea
\partial_\mu\varphi_c&=&0.
\label{eq:app_constant_background}
\eea
Under this condition, all terms containing derivatives of the background field vanish, and the effective action is determined by its zero-derivative sector. The explicit coordinate dependence of the corresponding local potential remains via the coefficient functions generated by the finite rotating geometry.

For the static and axially symmetric configurations considered in the main text, the explicit coordinate dependence reduces to the radial coordinate,
\bea
V_{\mathrm{eff}}(\varphi_c,x)&=&V_{\mathrm{eff}}(\varphi_c,r).
\label{eq:app_radial_effective_potential}
\eea
Within the local-potential approximation, the functional derivative of the effective action reduces at zeroth order in derivatives to
\bea
\frac{\delta\Gamma[\varphi_c]}{\delta\varphi_c(x)}&=&-\frac{\partial V_{\mathrm{eff}}(\varphi_c,r)}{\partial\varphi_c}.
\label{eq:app_local_functional_derivative}
\eea
Combining this relation with Eq.~\eqref{eq:effective_action_tadpole_relation} gives the one-loop relation
\bea
\frac{\partial V_{\mathrm{eff}}^{(1)}(\varphi_c,r)}{\partial\varphi_c}&=&\frac{\ii\lambda}{2}\varphi_c G_{\mathrm{F}}(x,x;\varphi_c).
\label{eq:effective_potential_derivative_GF}
\eea
For the constant background specified in Eq.~\eqref{eq:app_constant_background}, the background field is independent of $x$, whereas the coincident propagator retains the coordinate dependence associated with the finite rotating system.

The local extrema of the effective potential are consequently determined within this approximation by
\bea
\frac{\partial V_{\mathrm{eff}}(\varphi_c,r)}{\partial\varphi_c}&=&0.
\label{eq:app_local_stationarity}
\eea
The resulting solution $\varphi_c=\varphi_{\mathrm{ext}}(r)$ characterizes the extrema of the local effective potential at each radial position. This condition is distinct from the full stationarity condition obtained before imposing the local-potential approximation, which is given by
\bea
\frac{\delta\Gamma[\varphi_c]}{\delta\varphi_c(x)}&=&0.
\label{eq:app_full_stationarity}
\eea
The latter contains the contributions from the derivative sector and therefore determines the full background configuration when gradient terms are retained.

The spatially averaged effective potential can subsequently be introduced as
\bea
\overline{V}_{\mathrm{eff}}(\varphi_c)&=&\frac{1}{V}\int d^3x\,V_{\mathrm{eff}}(\varphi_c,r)\nn\\
&=&\frac{2}{R^2}\int_0^Rdr\,r\,V_{\mathrm{eff}}(\varphi_c,r),
\label{eq:app_volume_average}
\eea
where $V=\pi R^2L_z$ and the longitudinal dependence cancels for the static configurations considered here. Its stationary points satisfy
\bea
\frac{\partial\overline{V}_{\mathrm{eff}}}{\partial\varphi_c}&=&0.
\label{eq:app_global_stationarity}
\eea
The spatially resolved potential and its volume average therefore provide the local and spatially averaged descriptions of the effective potential, respectively.

As a consistency check, consider a background field with no spacetime dependence,
\bea
\partial_\mu\varphi_c&=&0.
\label{eq:app_constant_background_1}
\eea
In this limit, all derivative contributions vanish and Eq.~\eqref{eq:app_inhomogeneous_effective_action} reduces to
\bea
\Gamma[\varphi_c]&=&-\int d^4x\,V_{\mathrm{eff}}(\varphi_c),
\label{eq:app_homogeneous_effective_action}
\eea
which is the usual relation between the effective action and the effective potential for a homogeneous background. In the finite rotating system, the corresponding local quantities may retain an explicit radial dependence via the fluctuation spectrum even when the background field is constant.

The derivative expansion employed here is a bulk expansion in gradients of the background field. Since the finite cylinder introduces a boundary at $r=R$, the integrations by parts used above can generate surface contributions that are not contained in the bulk effective potential. These contributions are omitted in the present treatment. A systematic treatment of the boundary would require the inclusion of boundary-localized operators in the effective action. The Dirichlet condition imposed on the fluctuation modes determines the discrete Fourier-Bessel spectrum entering the propagator, while the local-potential approximation corresponds to the zero-derivative sector evaluated for a constant background field.

\subsection*{Thermal Propagator, Self-Energy, and One-Loop Effective Potential}

The one-loop effective potential is obtained from the relation between its derivative with respect to the constant background field and the coincident-point Feynman propagator derived in Eq.~\eqref{eq:effective_potential_derivative_GF}. The spatial mode decomposition required for the propagator was established in Eqs.~\eqref{eq:app_spatial_modes}--\eqref{eq:app_radial_normalization}, while the corresponding zero-temperature Feynman propagator was obtained in Eq.~\eqref{eq:app_scalar_propagator}. The finite-temperature calculation therefore starts directly from this propagator. For the constant background field introduced in Eq.~\eqref{eq:app_constant_background}, the effective mass and the corresponding mode frequency are
\bea
m_{\mathrm{eff}}^2(\varphi_c)&=&m^2+\frac{\lambda}{2}\varphi_c^2,\nn\\
\omega_{l,h,c}(k_z)&=&\sqrt{k_z^2+k_{\perp,l,h}^2+m_{\mathrm{eff}}^2(\varphi_c)}.
\label{eq:app_effective_mass_mode_energy}
\eea
The transverse momentum is fixed by the Dirichlet spectrum according to
\bea
k_{\perp,l,h}&=&\frac{\gamma_{l,h}}{R},
\label{eq:app_transverse_momentum}
\eea
where $\gamma_{l,h}$ denotes the $h$th positive zero of $J_l$. The quantity $\omega_{l,h,c}(k_z)$ is therefore the positive frequency associated with a fluctuation mode characterized by the quantum numbers $(l,h,k_z)$ in the presence of the constant background field.

We first consider the coincident-point limit of the zero-temperature propagator in Eq.~\eqref{eq:app_scalar_propagator}. For $x=(t,r,\phi,z)$ and $x'=(t',r',\phi',z')$, the limit $x'\rightarrow x$ implies $t'\rightarrow t$, $r'\rightarrow r$, $\phi'\rightarrow\phi$, and $z'\rightarrow z$. The phase factors in Eq.~\eqref{eq:app_scalar_propagator} consequently satisfy $e^{-\ii E(t-t')}\rightarrow1$, $e^{\ii k_z(z-z')}\rightarrow1$, and $e^{\ii l(\phi-\phi')}\rightarrow1$, while the product of radial mode functions becomes $J_l(k_{\perp,l,h}r)J_l(k_{\perp,l,h}r')\rightarrow[J_l(k_{\perp,l,h}r)]^2$. Hence, the coincidence limit removes the dependence on the coordinate differences $t-t'$, $z-z'$, and $\phi-\phi'$, but it does not remove the radial dependence. The two radial coordinates are simply identified, leaving the single coordinate $r$ at which the local propagator is evaluated. This residual dependence is a consequence of the finite cylindrical boundary, which breaks translational invariance in the transverse directions.

At finite temperature, the Euclidean time direction is compact with period $\beta=1/T$, and the continuous temporal frequency is replaced by the bosonic Matsubara frequencies
\bea
\omega_n&=&2\pi nT,\qquad n\in\mathbb{Z}.
\label{eq:app_matsubara_frequency}
\eea
The rotating dependence must be continued from the frequency combination appearing explicitly in Eq.~\eqref{eq:app_scalar_propagator}, namely $E+\Omega l$. Under the Wick rotation of the temporal frequency, $E\rightarrow\ii\omega_n$, this combination becomes
\bea
E+\Omega l&\longrightarrow&\ii\omega_n+\Omega l=\ii(\omega_n-\ii l\Omega).
\label{eq:app_rotating_matsubara_combination}
\eea
Consequently, the denominator of the propagator transforms according to
\bea
&&(E+\Omega l)^2-k_z^2-k_{\perp,l,h}^2-m_{\mathrm{eff}}^2(\varphi_c)+\ii\epsilon\nn\\
&\longrightarrow&-\left[(\omega_n-\ii l\Omega)^2+k_z^2+k_{\perp,l,h}^2+m_{\mathrm{eff}}^2(\varphi_c)\right]\nn\\
&=&-\left[(\omega_n-\ii l\Omega)^2+\omega_{l,h,c}^2(k_z)\right].
\label{eq:app_matsubara_denominator}
\eea
The overall sign generated by the Wick rotation is absorbed into the definition of the Euclidean propagator. Therefore, after taking the coincident-point limit and introducing the Matsubara frequencies, the finite-temperature propagator can be written as
\begin{widetext}
\bea
G_\beta(r,\Omega,T;\varphi_c)&=&\frac{T}{\pi R^2}\sum_{n=-\infty}^{\infty}\sum_{l=-\infty}^{\infty}\sum_{h=1}^{\infty}\frac{\left[J_l(k_{\perp,l,h}r)\right]^2}{\left[J_{l+1}(\gamma_{l,h})\right]^2}\int_{-\infty}^{\infty}\frac{dk_z}{2\pi}\frac{1}{(\omega_n-\ii l\Omega)^2+\omega_{l,h,c}^2(k_z)}.
\label{eq:app_thermal_propagator_matsubara}
\eea
\end{widetext}

The remaining frequency sum can be evaluated for each fixed set of quantum numbers $(l,h,k_z)$. Defining
\bea
S_\mathrm{E}(l,h,k_z)&=&T\sum_{n=-\infty}^{\infty}\frac{1}{(\omega_n-\ii l\Omega)^2+\omega_{l,h,c}^2(k_z)},…\nn\\
\label{eq:app_matsubara_sum}
\eea
the sum is evaluated using the Bose distribution
\bea
n_{\mathrm B}(z)&=&\frac{1}{e^{\beta z}-1}.
\label{eq:app_bose_distribution}
\eea

For the contour evaluation, it is convenient to introduce the complex frequency variable
\bea
z&=&\ii\omega_n+l\Omega.
\label{eq:app_matsubara_complex_variable}
\eea
With this definition, the denominator in Eq.~\eqref{eq:app_matsubara_sum} takes the form
\bea
(\omega_n-\ii l\Omega)^2+\omega_{l,h,c}^2&=&-\left[z^2-\omega_{l,h,c}^2\right].
\label{eq:app_matsubara_denominator_rewritten}
\eea
The propagator factor therefore has poles at
\bea
z_\pm&=&\pm\omega_{l,h,c}.
\label{eq:app_matsubara_poles}
\eea
The displacement of the Matsubara poles by $l\Omega$ is contained in the relation between $z$ and $\omega_n$, and consequently the Bose factors evaluated at the poles acquire the shifted arguments $\omega_{l,h,c}\pm l\Omega$. Evaluating the residues at the two poles yields
\bea
S_\mathrm{E}&=&\frac{1}{2\omega_{l,h,c}}\left[1+\sum_{s=\pm1}n_{\mathrm B}(\omega_{l,h,c}+sl\Omega)\right].
\label{eq:app_matsubara_result}
\eea
The first term in Eq.~\eqref{eq:app_matsubara_result} is independent of temperature and represents the zero-temperature contribution, while the two Bose distributions contain the finite-temperature correction and its dependence on the angular momentum $l$ and angular velocity $\Omega$. We therefore decompose the Matsubara sum as
\bea
S_\mathrm{E}&=&S_{\mathrm{vac}}+S_T,
\label{eq:app_vacuum_thermal_split}
\eea
where
\bea
S_{\mathrm{vac}}&=&\frac{1}{2\omega_{l,h,c}},\nn\\
S_T&=&\frac{n_{\mathrm B}(\omega_{l,h,c}-l\Omega)+n_{\mathrm B}(\omega_{l,h,c}+l\Omega)}{2\omega_{l,h,c}}.
\label{eq:app_vacuum_thermal_parts}
\eea
The vacuum term contains the ultraviolet divergence associated with the one-loop tadpole and is treated together with the vacuum counterterms according to the chosen renormalization prescription. The thermal term is ultraviolet finite and contains the explicit dependence on the temperature and angular velocity.

Substituting the thermal contribution in Eq.~\eqref{eq:app_vacuum_thermal_parts} into the mode expansion in Eq.~\eqref{eq:app_thermal_propagator_matsubara} gives the thermal part of the coincident-point propagator,
\begin{widetext}
\bea
G_T(r,\Omega,T;\varphi_c)&=&\frac{1}{\pi R^2}\sum_{l=-\infty}^{\infty}\sum_{h=1}^{\infty}\frac{\left[J_l(k_{\perp,l,h}r)\right]^2}{\left[J_{l+1}(\gamma_{l,h})\right]^2}\int_{-\infty}^{\infty}\frac{dk_z}{2\pi}\frac{n_{\mathrm B}\left(\omega_{l,h,c}(k_z)-l\Omega\right)+n_{\mathrm B}\left(\omega_{l,h,c}(k_z)+l\Omega\right)}{2\omega_{l,h,c}(k_z)}.
\label{eq:app_thermal_propagator}
\eea
\end{widetext}

The corresponding thermal contribution to the one-loop tadpole self-energy follows directly from the tadpole relation derived in Eq.~\eqref{eq:effective_action_tadpole_relation},
\bea
\Sigma_T(r,\Omega,T;\varphi_c)&=&\frac{\lambda}{2}G_T(r,\Omega,T;\varphi_c).
\label{eq:app_thermal_self_energy_relation}
\eea
The self-energy in Eq.~\eqref{eq:app_thermal_self_energy_relation} is local in the coordinate representation used in the local-potential approximation. Locality in coordinate space does not imply diagonality in the radial mode basis. Since $\Sigma_T$ depends on $r$, its projection onto the radial Bessel basis generally mixes different radial quantum numbers. This distinction is relevant for the ring resummation discussed in the subsequent appendix.

The Bose distributions appearing in Eq.~\eqref{eq:app_thermal_propagator} have positive arguments throughout the causal rotating region. The positive zeros of the Bessel functions satisfy
\bea
\gamma_{l,h}&>&|l|,
\label{eq:app_bessel_zero_bound}
\eea
which, together with Eq.~\eqref{eq:app_transverse_momentum}, implies
\bea
k_{\perp,l,h}&=&\frac{\gamma_{l,h}}{R}>\frac{|l|}{R}.
\label{eq:app_transverse_bound}
\eea
Using the causality condition in Eq.~\eqref{eq:rotating_causality_condition},
\bea
\Omega R&<&1,
\label{eq:app_rotational_causality}
\eea
we obtain
\bea
|l|\Omega&<&\frac{|l|}{R}<k_{\perp,l,h}\leq\omega_{l,h,c}(k_z).
\label{eq:app_rotational_bound}
\eea
Therefore,
\bea
\omega_{l,h,c}(k_z)\pm l\Omega&>&0.
\label{eq:app_positive_bose_arguments}
\eea
The thermal occupation factors are consequently finite for every allowed mode inside the causal region of the rotating cylinder.

We now return to the relation between the one-loop effective potential and the coincident-point propagator established in Eq.~\eqref{eq:effective_potential_derivative_GF}. Separating the propagator into vacuum and thermal contributions according to Eq.~\eqref{eq:app_vacuum_thermal_split} gives
\bea
\frac{\partial V_{\mathrm{eff}}^{(1)}(\varphi_c,r)}{\partial\varphi_c}&=&\frac{\partial V_{\mathrm{vac}}^{(1)}(\varphi_c,r)}{\partial\varphi_c}+\frac{\partial V_T^{(1)}(\varphi_c,r)}{\partial\varphi_c}.\nn\\
\label{eq:app_one_loop_derivative_split}
\eea
The thermal part consequently satisfies
\bea
\frac{\partial V_T^{(1)}(\varphi_c,r)}{\partial\varphi_c}&=&\frac{\lambda}{2}\varphi_cG_T(r,\Omega,T;\varphi_c).
\label{eq:app_effective_potential_derivative}
\eea
The background-field dependence enters through the effective mass in Eq.~\eqref{eq:app_effective_mass_mode_energy}. Differentiating the mode frequency with respect to $\varphi_c$ gives
\bea
\frac{\partial\omega_{l,h,c}(k_z)}{\partial\varphi_c}&=&\frac{\lambda\varphi_c}{2\omega_{l,h,c}(k_z)}.
\label{eq:app_mode_energy_derivative}
\eea
To integrate Eq.~\eqref{eq:app_effective_potential_derivative}, it is convenient to combine the two rotational contributions. For each value of $s$, the derivative of the corresponding thermal logarithm is
\bea
&&\frac{\partial}{\partial\varphi_c}\left\{T\ln\left[1-e^{-\beta(\omega_{l,h,c}(k_z)+sl\Omega)}\right]\right\}\nn\\
&=&\frac{\lambda\varphi_c}{2\omega_{l,h,c}(k_z)}n_{\mathrm B}\left(\omega_{l,h,c}(k_z)+sl\Omega\right),
\label{eq:app_thermal_log_derivative}
\eea
then, the sum over $s=\pm1$ therefore reproduces precisely the two Bose distributions appearing in Eq.~\eqref{eq:app_thermal_propagator}. Integration with respect to the background field then gives the thermal one-loop contribution, up to a function independent of $\varphi_c$,
\begin{widetext}
\bea
V_T^{(1)}(\varphi_c,r)&=&\frac{T}{2\pi R^2}\sum_{l=-\infty}^{\infty}\sum_{h=1}^{\infty}\frac{\left[J_l(k_{\perp,l,h}r)\right]^2}{\left[J_{l+1}(\gamma_{l,h})\right]^2}\int_{-\infty}^{\infty}\frac{dk_z}{2\pi}\sum_{s=\pm1}\ln\left[1-e^{-\beta(\omega_{l,h,c}(k_z)+sl\Omega)}\right]+C(r).
\label{eq:app_one_loop_thermal_potential}
\eea
\end{widetext}
where $C(r)$ is independent of $\varphi_c$. We choose the configuration $\varphi_c=0$ as the reference configuration, thereby fixing the field-independent normalization by requiring $V_T^{(1)}(0,r)=0$. The corresponding mode frequency is
\bea
\omega_{l,h,0}(k_z)&=&\sqrt{k_z^2+k_{\perp,l,h}^2+m^2}.
\label{eq:app_reference_mode_energy}
\eea
Subtracting the value at $\varphi_c=0$ gives
\begin{widetext}
\bea
V_T^{(1)}(\varphi_c,r)&=&\frac{T}{2\pi R^2}\sum_{l=-\infty}^{\infty}\sum_{h=1}^{\infty}\frac{\left[J_l(k_{\perp,l,h}r)\right]^2}{\left[J_{l+1}(\gamma_{l,h})\right]^2}\int_{-\infty}^{\infty}\frac{dk_z}{2\pi}\sum_{s=\pm1}\ln\left[\frac{1-e^{-\beta(\omega_{l,h,c}(k_z)+sl\Omega)}}{1-e^{-\beta(\omega_{l,h,0}(k_z)+sl\Omega)}}\right].
\label{eq:app_one_loop_thermal_potential_subtracted}
\eea
\end{widetext}
This subtraction fixes only the background-independent normalization of the thermal potential and does not modify its derivative with respect to $\varphi_c$.

The vacuum contribution follows from the first term in Eq.~\eqref{eq:app_vacuum_thermal_parts}. Applying Eq.~\eqref{eq:effective_potential_derivative_GF} to the vacuum contribution gives
\bea
\frac{\partial V_{\mathrm{vac}}^{(1)}(\varphi_c,r)}{\partial\varphi_c}&=&\frac{\lambda}{2}\varphi_cG_{\mathrm{vac}}(r;\varphi_c).
\label{eq:app_vacuum_potential_derivative}
\eea
Using the vacuum part of the same mode expansion gives
\bea
&&\frac{\partial V_{\mathrm{vac}}^{(1)}(\varphi_c,r)}{\partial\varphi_c}\nn\\
&=&\frac{\lambda\varphi_c}{4\pi R^2}\sum_{l=-\infty}^{\infty}\sum_{h=1}^{\infty}\frac{\left[J_l(k_{\perp,l,h}r)\right]^2}{\left[J_{l+1}(\gamma_{l,h})\right]^2}\int_{-\infty}^{\infty}\frac{dk_z}{2\pi}\frac{1}{\omega_{l,h,c}(k_z)}.\nn\\
\label{eq:app_vacuum_potential_derivative_explicit}
\eea

Using Eq.~\eqref{eq:app_mode_energy_derivative}, the integration with respect to $\varphi_c$ gives, up to a background-independent contribution,
\bea
&&V_{\mathrm{vac}}^{(1)}(\varphi_c,r)\nn\\
&=&\frac{1}{2\pi R^2}\sum_{l=-\infty}^{\infty}\sum_{h=1}^{\infty}\frac{\left[J_l(k_{\perp,l,h}r)\right]^2}{\left[J_{l+1}(\gamma_{l,h})\right]^2}\int_{-\infty}^{\infty}\frac{dk_z}{2\pi}\omega_{l,h,c}(k_z).\nn\\
\label{eq:app_one_loop_vacuum_potential}
\eea
This contribution is ultraviolet divergent and is understood together with the vacuum counterterms according to the chosen renormalization prescription. The transverse spectrum and radial mode functions depend on the cylinder radius and boundary condition, but not explicitly on $\Omega$. Consequently, the vacuum contribution has no explicit dependence on $\Omega$ in this representation.

Combining the vacuum and thermal contributions gives
\bea
V_{\mathrm{eff}}^{(1)}(\varphi_c,r)&=&V_{\mathrm{vac}}^{(1)}(\varphi_c,r)+V_T^{(1)}(\varphi_c,r).
\label{eq:app_one_loop_vacuum_thermal_split}
\eea
The classical contribution is
\bea
V^{(0)}(\varphi_c)&=&\frac{1}{2}m^2\varphi_c^2+\frac{\lambda}{4!}\varphi_c^4,
\label{eq:app_classical_potential}
\eea
and the effective potential in the local-potential approximation is therefore
\bea
V_{\mathrm{eff}}(\varphi_c,r)&=&V^{(0)}(\varphi_c)+V_{\mathrm{vac}}^{(1)}(\varphi_c,r)+V_T^{(1)}(\varphi_c,r).\nn\\
\label{eq:app_one_loop_effective_potential}
\eea

Substituting Eqs.~\eqref{eq:app_one_loop_vacuum_potential} and \eqref{eq:app_one_loop_thermal_potential_subtracted} gives the one-loop expression used in the main text,
\begin{widetext}
\bea
&&V_{\mathrm{eff}}(\varphi_c,r)\nn\\
&=&V^{(0)}(\varphi_c)+V_{\mathrm{vac}}^{(1)}(\varphi_c,r)+\frac{T}{2\pi R^2}\sum_{l=-\infty}^{\infty}\sum_{h=1}^{\infty}\frac{\left[J_l(k_{\perp,l,h}r)\right]^2}{\left[J_{l+1}(\gamma_{l,h})\right]^2}\int_{-\infty}^{\infty}\frac{dk_z}{2\pi}\sum_{s=\pm1}\ln\left[\frac{1-e^{-\beta(\omega_{l,h,c}(k_z)+sl\Omega)}}{1-e^{-\beta(\omega_{l,h,0}(k_z)+sl\Omega)}}\right].
\label{eq:app_final_one_loop_effective_potential}
\eea
\end{widetext}

\section{Functional Ring Resummation and Translational Symmetry Breaking}\label{app:ring-resummation}

This appendix presents the functional derivation of the ring resummation and its implementation in the rotating and spatially inhomogeneous system considered here. The derivation follows the standard reorganization of the perturbative expansion by adding and subtracting the thermal self-energy~\cite{Parwani1992}. This procedure incorporates the thermal self-energy into the quadratic fluctuation operator while retaining a compensating interaction that prevents double counting. We then formulate the resulting resummation at the operator level, which is necessary because the loss of transverse translational invariance makes the thermal self-energy position dependent and, in general, non-diagonal in the radial eigenmode basis.

At high temperature, the perturbative expansion becomes particularly sensitive to the infrared sector. In the vicinity of symmetry restoration, the effective classical mass can become small, $m_{\mathrm{eff}}^2\to0$, enhancing the contribution of the static Matsubara mode and producing infrared singularities in the conventional loop expansion~\cite{Parwani1992,kapusta2006finite,CurtinMeadeRamani2018}. The ring, or daisy, resummation reorganizes this expansion by incorporating the thermal self-energy into the propagator and thereby resumming the repeated self-energy insertions that become important in this regime.

In the rotating system considered here, the fluctuation operator is defined with respect to the Euclidean geometry and the boundary conditions associated with the rotating coordinates. We therefore formulate the resummation directly at the operator level, without assuming transverse translational invariance.

After Wick rotation to Euclidean spacetime, the quadratic action can be written by adding and subtracting the thermal self-energy $\Sigma_T(x)$:
\bea
&&S_\text{E}^{(2)}[\eta]\nn\\
&=&\frac{1}{2}\int d^4x_\text{E}\,\eta(x)\left[K_\text{E}(x;\varphi_c)+\Sigma_T(x)-\Sigma_T(x)\right]\eta(x).\nn\\
\label{eq:app_ring_quadratic_action}
\eea
Here, $K_\text{E}$ denotes the Euclidean fluctuation operator obtained from the original theory, including the dependence on the background field $\varphi_c$ and the rotational structure introduced by the chosen coordinates. In the absence of the thermal self-energy, its local mass contribution is given by $m^2+\frac{\lambda}{2}\varphi_c^2(x)$. The addition and subtraction of $\Sigma_T(x)$ does not modify the underlying theory, but reorganizes the perturbative expansion by defining an improved quadratic operator together with a compensating interaction:
\bea
S_\text{E}^{(2)}[\eta]&=&S_{\text{E},\mathrm{resum}}^{(2)}[\eta]+\Delta S_\text{E}[\eta],
\label{eq:app_ring_action_decomposition}
\eea
where
\bea
K_{\mathrm{resum}}(x;\varphi_c)&=&K_\text{E}(x;\varphi_c)+\Sigma_T(x),
\label{eq:app_ring_resummed_operator}
\eea
and
\bea
\Delta S_\text{E}[\eta]&=&-\frac{1}{2}\int d^4x_\text{E}\,\eta(x)\Sigma_T(x)\eta(x).
\label{eq:app_ring_counterterm_action}
\eea
The thermal self-energy is therefore included in the propagator used for the resummed expansion, while the subtraction term guarantees that the original quadratic theory is recovered when the expansion is carried out consistently. Since $\Sigma_T(x)$ is a local scalar function, it commutes with the scalar fluctuation field $\eta(x)$, and the subtraction term can equivalently be written as $-\frac{1}{2}\int d^4x_\text{E}\,\Sigma_T(x)\eta^2(x)$. Its nontrivial operator character arises when it is considered together with the differential operator $K_\text{E}$.

The corresponding Euclidean generating functional restricted to the quadratic fluctuations is
\bea
Z_\text{E}&\propto&\int\mathcal{D}\eta\,\exp\left\{-S_{\text{E},\mathrm{resum}}^{(2)}[\eta]-\Delta S_\text{E}[\eta]\right\}.
\label{eq:app_ring_generating_functional}
\eea
Treating the subtraction term as a perturbative insertion, its exponential can be expanded to first order:
\bea
Z_\text{E}&\simeq&\int\mathcal{D}\eta\,e^{-S_{\text{E},\mathrm{resum}}^{(2)}[\eta]}\left[1+\frac{1}{2}\int d^4x_\text{E}\,\eta(x)\Sigma_T(x)\eta(x)\right].\nn\\
\label{eq:app_ring_first_order_insertion}
\eea

The functional integration separates into the Gaussian determinant associated with the resummed operator and the Wick contraction generated by the subtraction term. Defining the resummed propagator as the inverse of the resummed fluctuation operator,
\bea
G_{\mathrm{resum}}(x,x';\varphi_c)&=&K_{\mathrm{resum}}^{-1}(x,x';\varphi_c),
\label{eq:app_ring_resummed_propagator}
\eea
the corresponding coincident contraction is
\bea
\frac{\int\mathcal{D}\eta\,\eta(x)\eta(x)e^{-S_{\text{E},\mathrm{resum}}^{(2)}}}{\int\mathcal{D}\eta\,e^{-S_{\text{E},\mathrm{resum}}^{(2)}}}&=&G_{\mathrm{resum}}(x,x;\varphi_c).
\label{eq:app_ring_coincident_contraction}
\eea
Then, the generating functional becomes
\bea
Z_\text{E}&\simeq&\left[\det K_{\mathrm{resum}}\right]^{-1/2}\nn\\
&\times&\left[1+\frac{1}{2}\int d^4x_\text{E}\,G_{\mathrm{resum}}(x,x;\varphi_c)\Sigma_T(x)\right].
\label{eq:app_ring_gaussian_result}
\eea
Taking the negative logarithm and retaining the contribution generated by the first-order subtraction insertion gives
\bea
\Gamma_{\text{E},\mathrm{resum}}&=&\frac{1}{2}\operatorname{Tr}\ln K_{\mathrm{resum}}-\frac{1}{2}\operatorname{Tr}\left(G_{\mathrm{resum}}\Sigma_T\right).\nn\\
\label{eq:app_ring_resummed_effective_action}
\eea
The first term contains the logarithmic determinant of the resummed operator and therefore generates the repeated thermal self-energy insertions. Its expansion also contains the contribution with a single insertion of $\Sigma_T$, which is already included in the ordinary one-loop expansion. To isolate the additional ring contribution and avoid double counting, we introduce the unperturbed thermal propagator $G_\beta\equiv K_\text{E}^{-1}$. The ring contribution can then be written as
\bea
\Delta\Gamma_{\mathrm{ring}}&=&\frac{1}{2}\operatorname{Tr}\ln\left(\mathbb{1}+G_\beta\Sigma_T\right)-\frac{1}{2}\operatorname{Tr}\left(G_\beta\Sigma_T\right).\nn\\
\label{eq:app_ring_operator_expression}
\eea
The logarithm generates the expansion
\bea
\operatorname{Tr}\ln\left(\mathbb{1}+G_\beta\Sigma_T\right)&=&\operatorname{Tr}(G_\beta\Sigma_T)-\frac{1}{2}\operatorname{Tr}(G_\beta\Sigma_TG_\beta\Sigma_T)\nn\\
&+&\frac{1}{3}\operatorname{Tr}(G_\beta\Sigma_TG_\beta\Sigma_TG_\beta\Sigma_T)-\cdots,\nn\\
\label{eq:app_ring_log_expansion}
\eea
so that the subtraction in Eq.~\eqref{eq:app_ring_operator_expression} removes the single-insertion contribution and retains the terms containing two or more thermal self-energy insertions. These terms constitute the ring, or daisy, contribution generated by the resummation.

In the present rotating system, the transverse translational invariance is broken by the finite cylindrical boundary. Consequently, $\Sigma_T(x)$ is generally position dependent and does not, in general, commute with the unperturbed fluctuation operator $K_\text{E}$. The thermal self-energy therefore cannot be treated, in general, as a simple mode-independent mass shift. Instead, its operator structure must be retained explicitly.

To express the logarithmic determinant in a form suitable for the inhomogeneous system, we introduce a continuous parameter $\alpha\in[0,1]$ and define the interpolating operator
\bea
K_{\beta,\alpha}&=&K_\text{E}+\alpha\Sigma_T.
\label{eq:app_ring_interpolating_operator}
\eea
The corresponding interpolated propagator is
\bea
G_{\beta,\alpha}&=&K_{\beta,\alpha}^{-1}.
\label{eq:app_ring_interpolating_propagator}
\eea
Using the operator identity under the functional trace,
\bea
&&\operatorname{Tr}\ln(K_\text{E}+\Sigma_T)-\operatorname{Tr}\ln K_\text{E}\nn\\
&=&\int_0^1d\alpha\,\operatorname{Tr}\left[(K_\text{E}+\alpha\Sigma_T)^{-1}\Sigma_T\right],
\label{eq:app_ring_trace_identity}
\eea
we obtain
\bea
\operatorname{Tr}\ln\left(\mathbb{1}+G_\beta\Sigma_T\right)&=&\int d^4x_\text{E}\int_0^1d\alpha\,G_{\beta,\alpha}(x,x;\varphi_c)\Sigma_T(x).\nn\\
\label{eq:app_ring_log_integral}
\eea
Since $G_{\beta,0}=G_\beta$, the subtraction of the single-insertion contribution can be incorporated into the same integral. The ring contribution therefore takes the form
\bea
&&\Delta\Gamma_{\mathrm{ring}}\nn\\
&=&\frac{1}{2}\int d^4x_\text{E}\int_0^1d\alpha\,\left[G_{\beta,\alpha}(x,x;\varphi_c)-G_\beta(x,x;\varphi_c)\right]\Sigma_T(x).\nn\\
\label{eq:app_ring_effective_action_alpha}
\eea
This expression remains valid without assuming that $\Sigma_T(x)$ is spatially constant or diagonal in the eigenmode basis of $K_\text{E}$. The dependence on the rotating geometry is contained in the Euclidean fluctuation operator, its spectrum, its eigenfunctions, and the corresponding coincident propagators.

Following the derivative expansion introduced in Eq.~\eqref{eq:app_homogeneous_effective_action}, the local zero-derivative contribution is identified with the ring correction to the effective potential:
\bea
&& V_{\mathrm{ring}}(\varphi_c,x)\nn\\
&=&\frac{1}{2}\int_0^1d\alpha\,\left[G_{\beta,\alpha}(x,x;\varphi_c)-G_\beta(x,x;\varphi_c)\right]\Sigma_T(\varphi_c,x).\nn\\
\label{eq:app_ring_local_potential}
\eea

The latter structure can be expressed formally by resolving the propagator and the thermal self-energy in the quantum numbers associated with the rotating cylindrical geometry. To represent the radial dependence, we use the normalized Bessel basis introduced in  Eq.~\eqref{eq:app_normalized_radial_modes} which satisfies the orthonormality relation given in Eq.~\eqref{eq:app_radial_normalization}. The resulting expression for the local ring contribution is
\bea
&& V_{\mathrm{ring}}(\varphi_c,r)\nn\\
&=&\frac{T}{2}\int_0^1d\alpha\sumint_{n,l,h,h'}dk_z\,U_{l,h}(r)\widetilde{\Sigma}^{(n,l,k_z)}_{h,h'} U_{l,h'}(r).\nn
\label{eq:app_ring_local_potential_matrix}
\eea
with
\bea
\widetilde{\Sigma}^{(n,l,k_z)}_{h,h'}\equiv\Big\{\left[G_{\beta,\alpha}^{(n,l,k_z)}\right]_{hh'}-\left[G_{\beta}^{(n,l,k_z)}\right]_{hh'}\Big\}\left[\Sigma_T^{(l)}\right]_{h'h}\nn\\
\eea
and
\bea
&&\sumint_{n,l,h,h'}dk_z\equiv\sum_{n=-\infty}^{\infty}\sum_{l=-\infty}^{\infty}\int_{-\infty}^{\infty}\frac{dk_z}{2\pi}\sum_{h,h'=1}^{\infty}.
\eea
Here, $n$, $l$, and $k_z$ label the Matsubara frequency, the azimuthal angular momentum, and the longitudinal momentum, respectively, while $h$ and $h'$ label the radial modes. The matrix structure in the radial indices arises precisely because the position dependence of $\Sigma_T(r,\Omega,T;\varphi_c)$ , determined by Eqs.~\eqref{eq:app_thermal_propagator} and \eqref{eq:app_thermal_self_energy_relation}, mixes different radial eigenmodes. In contrast, the absence of dependence on Euclidean time, the azimuthal angle, and the longitudinal coordinate ensures that no analogous mixing occurs among the corresponding quantum numbers.

The radial dependence of the thermal self-energy is represented by its matrix elements in the normalized basis introduced in Eq.~\eqref{eq:app_normalized_radial_modes}:
{\small
\bea
\left[\Sigma_T^{(l)}\right]_{hh'}&=&\int_0^Rr\,dr\,U_{l,h}(r)\Sigma_T(r,\Omega,T;\varphi_c)U_{l,h'}(r).\nn\\
&=&\frac{2}{R^2J_{l+1}(\gamma_{l,h})J_{l+1}(\gamma_{l,h'})}\nn\\
&\times&\int_0^Rr\,dr\,J_l(k_{\perp,l,h}r)\Sigma_T(r,\Omega,T;\varphi_c)J_l(k_{\perp,l,h'}r).\nn\\
\label{eq:app_ring_self_energy_bessel}
\eea
}

For fixed Matsubara frequency, angular momentum, and longitudinal momentum, the unperturbed Euclidean fluctuation operator is diagonal in the radial basis. Its matrix representation is
\bea
\left[K_{\beta}^{(n,l,k_z)}\right]_{hh'}&=&\delta_{hh'}\left[(\omega_n-\ii l\Omega)^2+\omega_{l,h,c}^2(k_z)\right],\nn\\
\label{eq:app_ring_unperturbed_operator}
\eea
where
\bea
\omega_{l,h,c}^2(k_z)&=&k_z^2+k_{\perp,l,h}^2+m_{\mathrm{eff}}^2(\varphi_c).
\label{eq:app_ring_mode_energy}
\eea
The interpolating operator defined in Eq.~\eqref{eq:app_ring_interpolating_operator} can consequently be represented in the radial mode space as
\bea
&&\left[K_{\beta,\alpha}^{(n,l,k_z)}\right]_{hh'}\nn\\
&=&\delta_{hh'}\left[(\omega_n-\ii l\Omega)^2+k_z^2+k_{\perp,l,h}^2+m_{\mathrm{eff}}^2(\varphi_c)\right]\nn\\
&+&\alpha\left[\Sigma_T^{(l)}\right]_{hh'}.
\label{eq:app_ring_interpolated_operator}
\eea
The corresponding propagator is obtained by matrix inversion in the radial indices:
\bea
\left[G_{\beta,\alpha}^{(n,l,k_z)}\right]_{hh'}&=&\left[\left(K_{\beta,\alpha}^{(n,l,k_z)}\right)^{-1}\right]_{hh'}.
\label{eq:app_ring_interpolated_propagator}
\eea
This matrix inverse is the explicit realization, in the radial mode basis, of the operator $G_{\beta,\alpha}$ entering Eq.~\eqref{eq:app_ring_effective_action_alpha}. The coincident-point propagator is then reconstructed from its radial matrix elements and the corresponding normalized radial functions:
\bea
G_{\beta,\alpha}(r,r;\varphi_c)&=&\frac{1}{\pi R^2}\sum_{l=-\infty}^{\infty}\sum_{h,h'=1}^{\infty}\frac{J_l(k_{\perp,l,h}r)J_l(k_{\perp,l,h'}r)}{J_{l+1}(\gamma_{l,h})J_{l+1}(\gamma_{l,h'})}\nn\\
&\times&T\sum_{n=-\infty}^{\infty}\int_{-\infty}^{\infty}\frac{dk_z}{2\pi}\left[G_{\beta,\alpha}^{(n,l,k_z)}\right]_{hh'}.
\label{eq:app_ring_coincident_propagator}
\eea

At $\alpha=0$, the interpolating operator reduces to the unresummed fluctuation operator and is diagonal in the radial basis. Its inverse is therefore
\bea
\left[G_{\beta,0}^{(n,l,k_z)}\right]_{hh'}&=&\frac{\delta_{hh'}}{(\omega_n-\ii l\Omega)^2+k_z^2+k_{\perp,l,h}^2+m_{\mathrm{eff}}^2(\varphi_c)}.\nn\\
\label{eq:app_ring_unperturbed_propagator}
\eea
Equations~\eqref{eq:app_ring_interpolated_operator} and \eqref{eq:app_ring_interpolated_propagator} show explicitly how the position-dependent thermal self-energy modifies the radial fluctuation spectrum. For $\alpha\neq0$, the off-diagonal elements of $\Sigma_T^{(l)}$ couple different radial modes, and the resummed propagator must therefore be obtained from the full matrix inverse rather than from independent mode-by-mode shifts.

The Matsubara summation can be performed by resolving the radial structure of the interpolating propagator spectrally. For fixed $l$, $\alpha$, and $k_z$, we define the radial operator
{\small
\bea
\left[\mathcal{M}_{l,\alpha}(k_z)\right]_{hh'}&=&\delta_{hh'}\left[k_z^2+k_{\perp,l,h}^2+m_{\mathrm{eff}}^2(\varphi_c)\right]+\alpha\left[\Sigma_T^{(l)}\right]_{hh'},\nn\\
\label{eq:app_ring_radial_operator}
\eea
}%
whose eigenvalues and normalized eigenvectors are defined by
\bea
\sum_{h'=1}^{\infty}\left[\mathcal{M}_{l,\alpha}(k_z)\right]_{hh'}u_{l,a}^{(\alpha)}(h';k_z)&=&\lambda_{l,a}^{(\alpha)}(k_z)u_{l,a}^{(\alpha)}(h;k_z),\nn\\
\label{eq:app_ring_radial_eigenvalue_problem}
\eea
with
\bea
\sum_{h=1}^{\infty}u_{l,a}^{(\alpha)*}(h;k_z)u_{l,b}^{(\alpha)}(h;k_z)&=&\delta_{ab}.
\label{eq:app_ring_eigenvector_orthonormality}
\eea
The interpolating operator has the same eigenvectors as the radial operator. Indeed, using Eq.~\eqref{eq:app_ring_radial_eigenvalue_problem}, its action on an eigenvector gives
{\small
\bea
K_{\beta,\alpha}^{(n,l,k_z)}u_{l,a}^{(\alpha)}(h;k_z)&=&\left[(\omega_n-\ii l\Omega)^2+\lambda_{l,a}^{(\alpha)}(k_z)\right]u_{l,a}^{(\alpha)}(h;k_z).\nn\\
\label{eq:app_ring_interpolated_eigenvalue}
\eea
}
Thus, in the spectral basis, the inverse operator is obtained by inverting the corresponding eigenvalues. Using the orthonormality of the eigenvectors, the propagator in Eq.~\eqref{eq:app_ring_interpolated_propagator} becomes
\bea
\left[G_{\beta,\alpha}^{(n,l,k_z)}\right]_{hh'}&=&\sum_{a=1}^{\infty}\frac{u_{l,a}^{(\alpha)}(h;k_z)u_{l,a}^{(\alpha)*}(h';k_z)}{(\omega_n-\ii l\Omega)^2+\lambda_{l,a}^{(\alpha)}(k_z)}.
\label{eq:app_ring_spectral_propagator}
\eea

The Matsubara summation can therefore be performed for each eigenvalue. Using the rotational shift $\omega_n\rightarrow\omega_n-\ii l\Omega$, one obtains
\bea
&&T\sum_{n=-\infty}^{\infty}\left[G_{\beta,\alpha}^{(n,l,k_z)}\right]_{hh'}\nn\\
&=&\sum_{a=1}^{\infty}u_{l,a}^{(\alpha)}(h;k_z)u_{l,a}^{(\alpha)*}(h';k_z)\nn\\
&\times&\frac{1}{2\sqrt{\lambda_{l,a}^{(\alpha)}(k_z)}}\left[1+\sum_{s=\pm1}n_\text{B}\left(\sqrt{\lambda_{l,a}^{(\alpha)}(k_z)}+sl\Omega\right)\right],\nn\\
\label{eq:app_ring_matsubara_sum}
\eea
where
\bea
n_\text{B}(E)&\equiv&\frac{1}{e^{\beta E}-1}.
\label{eq:app_ring_bose_distribution}
\eea

For $\alpha=0$, the radial operator is diagonal in the Bessel basis. Its eigenvalues and eigenvectors are therefore
\bea
\lambda_{l,h}^{(0)}(k_z)&=&k_z^2+k_{\perp,l,h}^2+m_{\mathrm{eff}}^2(\varphi_c),
\label{eq:app_ring_unperturbed_eigenvalues}
\eea
and
\bea
u_{l,a}^{(0)}(h;k_z)&=&\delta_{ah}.
\label{eq:app_ring_unperturbed_eigenvectors}
\eea
The corresponding Matsubara sum follows from Eq.~\eqref{eq:app_ring_matsubara_sum} by setting $\alpha=0$. 

Substituting the resummed and unperturbed Matsubara sums into the local ring contribution in Eq.~\eqref{eq:app_ring_local_potential_matrix}, and using the Bessel representation of the thermal self-energy matrix, the full finite-temperature ring correction becomes
\begin{widetext}
   \bea
   && V_{\mathrm{ring}}(\varphi_c,r)\nn\\
&=&\frac{1}{R^4}\int_0^1d\alpha\sum_{l=-\infty}^{\infty}\sum_{h,h'=1}^{\infty}\frac{J_l(k_{\perp,l,h}r)J_l(k_{\perp,l,h'}r)}{J_{l+1}(\gamma_{l,h})J_{l+1}(\gamma_{l,h'})}\int_0^Rr'dr'\frac{J_l(k_{\perp,l,h'}r')J_l(k_{\perp,l,h}r')}{J_{l+1}(\gamma_{l,h'})J_{l+1}(\gamma_{l,h})}\Sigma_T(r',\Omega,T;\varphi_c)\nn\\
&\times&\int_{-\infty}^{\infty}\frac{dk_z}{2\pi}\left\{\sum_{a=1}^{\infty}u_{l,a}^{(\alpha)}(h;k_z)u_{l,a}^{(\alpha)*}(h';k_z)\frac{1}{2\sqrt{\lambda_{l,a}^{(\alpha)}(k_z)}}\left[1+\sum_{s=\pm1}n_\text{B}\left(\sqrt{\lambda_{l,a}^{(\alpha)}(k_z)}+sl\Omega\right)\right]\right.\nn\\
&&\left.-\delta_{hh'}\frac{1}{2\sqrt{\lambda_{l,h}^{(0)}(k_z)}}\left[1+\sum_{s=\pm1}n_\text{B}\left(\sqrt{\lambda_{l,h}^{(0)}(k_z)}+sl\Omega\right)\right]\right\}.
\label{eq:app_ring_finiteT_local_potential_full}
\eea
Since the self-energy entering the ring resummation is defined as the thermal tadpole contribution in Eq.~\eqref{eq:app_thermal_self_energy_relation}, we retain only the thermal part of the Matsubara sums in the ring correction. The vacuum contributions are therefore omitted, and we define the thermal ring potential as
\bea
&& V_{\mathrm{ring}}^{T}(\varphi_c,r)\nn\\
&=&\frac{1}{R^4}\int_0^1d\alpha\sum_{l=-\infty}^{\infty}\sum_{h,h'=1}^{\infty}\frac{J_l(k_{\perp,l,h}r)J_l(k_{\perp,l,h'}r)}{J_{l+1}(\gamma_{l,h})J_{l+1}(\gamma_{l,h'})}\int_0^Rr'dr'\frac{J_l(k_{\perp,l,h'}r')J_l(k_{\perp,l,h}r')}{J_{l+1}(\gamma_{l,h'})J_{l+1}(\gamma_{l,h})}\Sigma_T(r',\Omega,T;\varphi_c)\nn\\
&&\times\int_{-\infty}^{\infty}\frac{dk_z}{2\pi}\left\{\sum_{a=1}^{\infty}u_{l,a}^{(\alpha)}(h;k_z)u_{l,a}^{(\alpha)*}(h';k_z)\frac{1}{2\sqrt{\lambda_{l,a}^{(\alpha)}(k_z)}}\sum_{s=\pm1}n_\text{B}\left(\sqrt{\lambda_{l,a}^{(\alpha)}(k_z)}+sl\Omega\right)\right.\nn\\
&&\left.-\delta_{hh'}\frac{1}{2\sqrt{\lambda_{l,h}^{(0)}(k_z)}}\sum_{s=\pm1}n_\text{B}\left(\sqrt{\lambda_{l,h}^{(0)}(k_z)}+sl\Omega\right)\right\}.
\label{eq:app_ring_finiteT_local_potential_thermal}
\eea
\end{widetext}
In what follows, $ V_{\mathrm{ring}}^{T}(\varphi_c,r)$ denotes the thermal ring contribution to the effective potential.

\section{High-Temperature Expansion and Macroscopic Analytical Simplification}
\label{app:high_temp_expansion}

To analyze the macroscopic phase structure of the system, it is useful to consider the high-temperature limit of the thermal contribution to the effective potential. The thermal ring contribution derived in Appendix~\ref{app:ring-resummation} is expressed after the thermal frequency sum in terms of Bose-Einstein distribution functions. The high-temperature approximation can therefore be implemented directly by expanding these distribution functions for energies small compared with the temperature.

For the thermal ring contribution, the relevant factor appearing in Eq.~\eqref{eq:app_ring_finiteT_local_potential_thermal} is
\bea
\frac{1}{2\sqrt{\lambda_{l,a}^{(\alpha)}(k_z)}}\sum_{s=\pm1}n_{\mathrm B}\left(\sqrt{\lambda_{l,a}^{(\alpha)}(k_z)}+sl\Omega\right).
\label{eq:app_HTE_thermal_factor}
\eea
In the high-temperature regime, the Bose-Einstein distribution can be expanded as
\bea
n_{\mathrm B}(E)&\simeq&\frac{T}{E}.
\label{eq:app_HTE_bose_expansion}
\eea
Applying this approximation to Eq.~\eqref{eq:app_HTE_thermal_factor} gives
\bea
&&\frac{1}{2\sqrt{\lambda_{l,a}^{(\alpha)}(k_z)}}\sum_{s=\pm1}n_{\mathrm B}\left(\sqrt{\lambda_{l,a}^{(\alpha)}(k_z)}+sl\Omega\right)\nn\\
&\simeq&\frac{T}{2\sqrt{\lambda_{l,a}^{(\alpha)}(k_z)}}\sum_{s=\pm1}\frac{1}{\sqrt{\lambda_{l,a}^{(\alpha)}(k_z)}+sl\Omega}\nn\\
&=&\frac{T}{\lambda_{l,a}^{(\alpha)}(k_z)-l^2\Omega^2}.
\label{eq:app_HTE_bose_reduction}
\eea
Thus, rotation enters the leading high-temperature contribution through the combination $\lambda_{l,a}^{(\alpha)}(k_z)-l^2\Omega^2$.

For fixed $l$ and $\alpha$, the eigenvalues of the radial operator introduced in Eq.~\eqref{eq:app_ring_radial_operator} can be written as
\bea
\lambda_{l,a}^{(\alpha)}(k_z)&=&k_z^2+\lambda_{l,a}^\perp(\alpha).
\label{eq:app_HTE_spectral_decomposition}
\eea
where $\lambda_{l,a}^\perp(\alpha)$ denotes the eigenvalues of the transverse radial operator before the rotational shift. We therefore define
\bea
M_{l,a}^2(\alpha)&\equiv&\lambda_{l,a}^\perp(\alpha)-l^2\Omega^2,
\label{eq:app_HTE_mass_definition}
\eea
so that
\bea
\lambda_{l,a}^{(\alpha)}(k_z)-l^2\Omega^2&=&k_z^2+M_{l,a}^2(\alpha).
\label{eq:app_HTE_longitudinal_denominator}
\eea
The longitudinal momentum integration can then be performed analytically whenever $M_{l,a}^2(\alpha)>0$,
\bea
\int_{-\infty}^{\infty}\frac{dk_z}{2\pi}\frac{T}{k_z^2+M_{l,a}^2(\alpha)}&=&\frac{T}{2M_{l,a}(\alpha)}.
\label{eq:app_HTE_kz_integral}
\eea
At the operator level, if $\mathcal M_{l,\alpha}$ denotes the transverse operator whose eigenvalues are $M_{l,a}^2(\alpha)$, this relation becomes
\bea
\int_{-\infty}^{\infty}\frac{dk_z}{2\pi}T\left[k_z^2\mathbb{1}+\mathcal M_{l,\alpha}\right]^{-1}&=&\frac{T}{2}\mathcal M_{l,\alpha}^{-1/2}.
\label{eq:app_HTE_operator_integral}
\eea
The continuous longitudinal momentum integration is therefore reduced to a matrix function of the transverse operator.

The remaining spatial dependence is treated by averaging the local ring contribution over the transverse area,
\bea
\overline{V}_{\mathrm{ring}}^{\mathrm{HT}}(\varphi_c)&=&\frac{2}{R^2}\int_0^Rdr\,r\,V_{\mathrm{ring}}^{\mathrm{HT}}(\varphi_c,r).
\label{eq:app_HTE_volume_average}
\eea
Using the normalized radial modes introduced in Eq.~\eqref{eq:app_normalized_radial_modes}, their orthonormality relation gives
\bea
\int_0^Rdr\,r\,U_{l,h}(r)U_{l,h'}(r)&=&\delta_{hh'}.
\label{eq:app_HTE_normalized_orthogonality}
\eea
or, equivalently, in terms of the Bessel functions,
\bea
\int_0^Rdr\,r\,J_l(k_{\perp,l,h}r)J_l(k_{\perp,l,h'}r)&=&\frac{R^2}{2}J_{l+1}^2(\gamma_{l,h})\delta_{hh'}.\nn\\
\label{eq:app_HTE_bessel_orthogonality}
\eea
The spatial average therefore produces a trace over the radial mode space. This operation does not remove the radial mode mixing generated by the position-dependent thermal self-energy, so the off-diagonal elements of $\Sigma_T^{(l)}$ remain in the corresponding matrix products. For a radial operator $\mathcal O$, the trace is defined by
\bea
\sum_{h,h'=1}^{\infty}\delta_{hh'}\mathcal O_{h'h}&=&\sum_{h=1}^{\infty}\mathcal O_{hh}\equiv\operatorname{Tr}_h(\mathcal O).
\label{eq:app_HTE_trace_definition}
\eea
Using the matrix representation of the thermal self-energy and Eq.~\eqref{eq:app_HTE_operator_integral}, the volume-averaged high-temperature ring contribution becomes
\bea
&&\overline{V}_{\mathrm{ring}}^{\mathrm{HT}}(\varphi_c)\nn\\
&=&\frac{T}{4\pi R^2}\sum_{l=-\infty}^{\infty}\int_0^1d\alpha\,\operatorname{Tr}_h\left\{\Sigma_T^{(l)}\left[\mathcal M_{l,\alpha}^{-1/2}-\mathcal M_{l,0}^{-1/2}\right]\right\}.\nn\\
\label{eq:app_HTE_volume_averaged_ring}
\eea
The interpolating operator satisfies
\bea
\mathcal M_{l,\alpha}&=&\mathcal M_{l,0}+\alpha\Sigma_T^{(l)},
\label{eq:app_HTE_interpolating_operator}
\eea
and therefore
\bea
\frac{d\mathcal M_{l,\alpha}}{d\alpha}&=&\Sigma_T^{(l)}.
\label{eq:app_HTE_alpha_derivative}
\eea
For a positive operator, the derivative of the trace of its square root is
\bea
\frac{d}{d\alpha}\operatorname{Tr}_h\left(\mathcal M_{l,\alpha}^{1/2}\right)&=&\frac{1}{2}\operatorname{Tr}_h\left(\Sigma_T^{(l)}\mathcal M_{l,\alpha}^{-1/2}\right).
\label{eq:app_HTE_trace_derivative}
\eea
Therefore, the auxiliary-parameter integration gives
\bea
\overline{V}_{\mathrm{ring}}^{\mathrm{HT}}(\varphi_c)&=&\frac{T}{2\pi R^2}\sum_{l=-\infty}^{\infty}\Big\{\operatorname{Tr}_h\left(\mathcal M_{l,1}^{1/2}-\mathcal M_{l,0}^{1/2}\right)\nn\\
&-&\frac{1}{2}\operatorname{Tr}_h\left(\Sigma_T^{(l)}\mathcal M_{l,0}^{-1/2}\right)\Big\}.
\label{eq:app_HTE_integrated_ring}
\eea
The one-loop thermal contribution is treated with the same leading high-temperature approximation. Its infrared-sensitive part can be written in terms of the unresummed transverse operator as
\bea
\overline{V}_{T,\mathrm{IR}}^{(1)}(\varphi_c)&=&\frac{T}{2\pi R^2}\sum_{l=-\infty}^{\infty}\operatorname{Tr}_h\int_{-\infty}^{\infty}\frac{dk_z}{2\pi}\ln\left[k_z^2\mathbb{1}+\mathcal M_{l,0}\right].\nn\\
\label{eq:app_HTE_oneloop_integral}
\eea
The longitudinal integral satisfies
\bea
\int_{-\infty}^{\infty}\frac{dk_z}{2\pi}\ln\left(k_z^2\mathbb{1}+\mathcal M_{l,0}\right)&=&\mathcal M_{l,0}^{1/2}+C,
\label{eq:app_HTE_dimreg_identity}
\eea
where $C$ is independent of the background field. Hence,
\bea
\overline{V}_{T,\mathrm{IR}}^{(1)}(\varphi_c)&=&\frac{T}{2\pi R^2}\sum_{l=-\infty}^{\infty}\operatorname{Tr}_h\left(\mathcal M_{l,0}^{1/2}\right)+C(T,\Omega,R).\nn\\
\label{eq:app_HTE_oneloop_final}
\eea
Combining this contribution with Eq.~\eqref{eq:app_HTE_integrated_ring}, the terms involving $\mathcal M_{l,0}^{1/2}$ cancel, giving
\bea
&&\overline{V}_{T,\mathrm{IR}}^{(1)}+\overline{V}_{\mathrm{ring}}^{\mathrm{HT}}\nn\\
&=&\frac{T}{2\pi R^2}\sum_{l=-\infty}^{\infty}\left[\operatorname{Tr}_h\left(\mathcal M_{l,1}^{1/2}\right)-\frac{1}{2}\operatorname{Tr}_h\left(\Sigma_T^{(l)}\mathcal M_{l,0}^{-1/2}\right)\right]\nn\\
&+&C(T,\Omega,R).
\label{eq:app_HTE_resummed_infrared}
\eea
Within the leading high-temperature approximation, the resulting macroscopic effective potential is therefore
\bea
\overline{V}_{\mathrm{eff}}^{\mathrm{HT}}(\varphi_c)&=&\frac{1}{2}m^2\varphi_c^2+\frac{\lambda}{4!}\varphi_c^4+\overline{V}_{\mathrm{vac}}^{(1)}(\varphi_c)\nn\\
&&+\frac{T}{2\pi R^2}\sum_{l=-\infty}^{\infty}\Big[\operatorname{Tr}_h\left(\mathcal M_{l,1}^{1/2}(\varphi_c)\right)\nn\\
&-&\frac{1}{2}\operatorname{Tr}_h\left(\Sigma_T^{(l)}(\varphi_c)\mathcal M_{l,0}^{-1/2}(\varphi_c)\right)\Big].
\label{eq:app_HTE_final_computational}
\eea
Here, the rotational dependence is contained in the transverse eigenvalues $M_{l,a}^2(\alpha)=\lambda_{l,a}^\perp(\alpha)-l^2\Omega^2$, while the finite transverse size enters through the discrete radial spectrum.

\bibliography{ScalarPropagatorLarry.bib}

@article{becattini2015vorticity,
  author        = {Becattini, F. and Csernai, L. P. and Wang, D. J.},
  title         = {Vorticity formation in high energy nuclear collisions},
  journal       = {Phys. Rev. C},
  volume        = {93},
  pages         = {024902},
  year          = {2016},
  doi           = {10.1103/PhysRevC.93.024902},
  eprint        = {1502.00386},
  archivePrefix = {arXiv},
  primaryClass  = {nucl-th}
}

@article{CurtinMeadeRamani2018,
  author  = {Curtin, David and Meade, Patrick and Ramani, Chanda},
  title   = {Thermal Resummation and Phase Transitions},
  journal = {Eur. Phys. J. C},
  volume  = {78},
  pages   = {787},
  year    = {2018},
  doi     = {10.1140/epjc/s10052-018-6268-0}
}

@article{Parwani1992,
  author  = {Parwani, Rajesh R.},
  title   = {Resummation in a hot scalar field theory},
  journal = {Phys. Rev. D},
  volume  = {45},
  pages   = {4695--4705},
  year    = {1992},
  doi     = {10.1103/PhysRevD.45.4695}
}

@article{deng2016vorticity,
  author        = {Deng, X.-G. and Huang, X.-G.},
  title         = {Vorticity in Heavy-Ion Collisions},
  journal       = {Phys. Rev. C},
  volume        = {93},
  pages         = {064907},
  year          = {2016},
  doi           = {10.1103/PhysRevC.93.064907},
  eprint        = {1603.06117},
  archivePrefix = {arXiv},
  primaryClass  = {nucl-th}
}

@article{ivanov2020vorticity,
  author        = {Ivanov, E. V. and Toneev, V. D. and Soldatov, A. A.},
  title         = {Vorticity and polarization in heavy-ion collisions},
  journal       = {Phys. Atom. Nucl.},
  volume        = {83},
  pages         = {179},
  year          = {2020},
  doi           = {10.1134/S1063778820020128},
  eprint        = {1910.01332},
  archivePrefix = {arXiv},
  primaryClass  = {nucl-th}
}

@article{star2017polarization,
  author        = {Adam, J. and others},
  collaboration = {STAR},
  title         = {Global $\Lambda$ hyperon polarization in nuclear collisions: evidence for the most vortical fluid},
  journal       = {Nature},
  volume        = {548},
  pages         = {62--65},
  year          = {2017},
  doi           = {10.1038/nature23004},
  eprint        = {1701.06657},
  archivePrefix = {arXiv},
  primaryClass  = {nucl-ex}
}

@article{fukushima2019extreme,
  author        = {Fukushima, K.},
  title         = {Extreme matter in electromagnetic fields and rotation},
  journal       = {Prog. Part. Nucl. Phys.},
  volume        = {107},
  pages         = {167--199},
  year          = {2019},
  doi           = {10.1016/j.ppnp.2019.04.001},
  eprint        = {1812.08886},
  archivePrefix = {arXiv},
  primaryClass  = {hep-ph}
}

@article{vilenkin1980rotation,
  author  = {Vilenkin, A.},
  title   = {Quantum field theory at finite temperature in a rotating system},
  journal = {Phys. Rev. D},
  volume  = {21},
  pages   = {2260--2269},
  year    = {1980},
  doi     = {10.1103/PhysRevD.21.2260}
}

@article{kuboniwa2026perturbation,
  author        = {Kuboniwa, Ryo and Mameda, Kazuya},
  title         = {Perturbation theory of rotating scalar fields and vacuum insensitivity to rotation},
  journal       = {Phys. Lett. B},
  volume        = {872},
  pages         = {140089},
  year          = {2026},
  doi           = {10.1016/j.physletb.2025.140089},
  eprint        = {2504.04712},
  archivePrefix = {arXiv},
  primaryClass  = {hep-th}
}

@article{kawaguchi2025susceptibilities,
  author        = {Kawaguchi, M. and Mameda, K.},
  title         = {Rotating finite-size systems and susceptibilities in the Nambu--Jona-Lasinio model},
  journal       = {JHEP},
  volume        = {11},
  pages         = {170},
  year          = {2025},
  doi           = {10.1007/JHEP11(2025)170},
  eprint        = {2507.00494},
  archivePrefix = {arXiv},
  primaryClass  = {hep-ph}
}

@article{salvio2026scalar,
  author        = {Salvio, Alberto},
  title         = {Scalar thermal field theory for a rotating plasma},
  journal       = {Eur. Phys. J. C},
  volume        = {86},
  pages         = {436},
  year          = {2026},
  doi           = {10.1140/epjc/s10052-026-15465-9},
  eprint        = {2503.09677},
  archivePrefix = {arXiv},
  primaryClass  = {hep-ph}
}

@article{PhysRevD.108.085016,
  title = {Rigidly rotating scalar fields: Between real divergence and imaginary fractalization},
  author = {Ambru{\ifmmode \mbox{\c{s}}\else \c{s}\fi{}}, Victor E. and Chernodub, Maxim N.},
  journal = {Phys. Rev. D},
  volume = {108},
  issue = {8},
  pages = {085016},
  numpages = {42},
  year = {2023},
  month = {Oct},
  publisher = {American Physical Society},
  doi = {10.1103/PhysRevD.108.085016},
  url = {https://link.aps.org/doi/10.1103/PhysRevD.108.085016}
}

@article{PhysRevD.110.094053, title = {Inhibition of the splitting of the chiral and deconfinement transition due to rotation in QCD: The phase diagram of the linear sigma model coupled to Polyakov loops}, author = {Singha, Pracheta and Ambru{\ifmmode \mbox{\c{s}}\else \c{s}\fi{}}, Victor E. and Chernodub, Maxim N.}, journal = {Phys. Rev. D}, volume = {110}, issue = {9}, pages = {094053}, numpages = {18}, year = {2024}, month = {Nov}, publisher = {American Physical Society}, doi = {10.1103/PhysRevD.110.094053}, url = {https://link.aps.org/doi/10.1103/PhysRevD.110.094053} }

@article{4zrn-wgrg, title = {Firewall boundaries and mixed phases of rotating quark matter in linear sigma model}, author = {Morales Tejera, Sergio and Ambru{\ifmmode \mbox{\c{s}}\else \c{s}\fi{}}, Victor E. and Chernodub, Maxim N.}, journal = {Phys. Rev. D}, volume = {112}, issue = {5}, pages = {054031}, numpages = {25}, year = {2025}, month = {Sep}, publisher = {American Physical Society}, doi = {10.1103/4zrn-wgrg}, url = {https://link.aps.org/doi/10.1103/4zrn-wgrg} }

@article{knn8-sv3k, title = {Linear sigma model with quarks and Polyakov loop in rotation: Phase diagrams, Tolman-Ehrenfest law, and mechanical properties}, author = {Singha, Pracheta and Busuioc, Sergiu and Ambru{\ifmmode \mbox{\c{s}}\else \c{s}\fi{}}, Victor E. and Chernodub, Maxim N.}, journal = {Phys. Rev. D}, volume = {112}, issue = {9}, pages = {094031}, numpages = {40}, year = {2025}, month = {Nov}, publisher = {American Physical Society}, doi = {10.1103/knn8-sv3k}, url = {https://link.aps.org/doi/10.1103/knn8-sv3k} }

@article{PhysRevD.103.094515, title = {Influence of relativistic rotation on the confinement-deconfinement transition in gluodynamics}, author = {Braguta, V. V. and Kotov, A. Yu. and Kuznedelev, D. D. and Roenko, A. A.}, journal = {Phys. Rev. D}, volume = {103}, issue = {9}, pages = {094515}, numpages = {16}, year = {2021}, month = {May}, publisher = {American Physical Society}, doi = {10.1103/PhysRevD.103.094515}, url = {https://link.aps.org/doi/10.1103/PhysRevD.103.094515} }

@article{dolan1974symmetry,
  author  = {Dolan, L. and Jackiw, R.},
  title   = {Symmetry behavior at finite temperature},
  journal = {Phys. Rev. D},
  volume  = {9},
  pages   = {3320--3341},
  year    = {1974},
  doi     = {10.1103/PhysRevD.9.3320}
}

@article{gaspar2023chiral,
  author        = {Gaspar, I. and Hern{\'a}ndez, A. and Zamora, R.},
  title         = {Chiral symmetry restoration in a rotating Yukawa model},
  journal       = {Phys. Rev. D},
  volume        = {108},
  pages         = {094020},
  year          = {2023},
  doi           = {10.1103/PhysRevD.108.094020},
  eprint        = {2305.00101},
  archivePrefix = {arXiv},
  primaryClass  = {hep-ph}
}

@article{hernandez2025vortical,
  author        = {Hern{\'a}ndez, A. and Zamora, R.},
  title         = {Vortical effects on the chiral phase transition and critical end point},
  journal       = {Phys. Rev. D},
  volume        = {111},
  pages         = {036003},
  year          = {2025},
  doi           = {10.1103/PhysRevD.111.036003},
  eprint        = {2410.17874},
  archivePrefix = {arXiv},
  primaryClass  = {hep-ph}
}

@article{siri2024bose,
  author        = {Siri, A. and Sadooghi, N.},
  title         = {Rigidly rotating relativistic Bose gas},
  journal       = {Phys. Rev. D},
  volume        = {110},
  pages         = {036016},
  year          = {2024},
  doi           = {10.1103/PhysRevD.110.036016},
  eprint        = {2405.09481},
  archivePrefix = {arXiv},
  primaryClass  = {hep-ph}
}

@article{PhysRevD.110.056014,
  title = {Temperature fluctuations in a relativistic gas: Pressure corrections and possible consequences in the deconfinement transition},
  author = {Casta{\~n}o-Yepes, Jorge David and Loewe, Marcelo and Mu{\~n}oz, Enrique and Rojas, Juan Crist{\'o}bal},
  journal = {Phys. Rev. D},
  volume = {110},
  issue = {5},
  pages = {056014},
  numpages = {12},
  year = {2024},
  month = {Sep},
  publisher = {American Physical Society},
  doi = {10.1103/PhysRevD.110.056014},
  url = {https://link.aps.org/doi/10.1103/PhysRevD.110.056014}
}

@article{PhysRevD.106.116019,
  title = {Volume effects on the QCD critical end point from thermal fluctuations within the super statistics framework},
  author = {Casta{\~n}o-Yepes, Jorge David and Mart{\'\i}nez Paniagua, Fernando and Mu{\~n}oz-Vitelly, Victor and Ramirez-Gutierrez, Cristian Felipe},
  journal = {Phys. Rev. D},
  volume = {106},
  issue = {11},
  pages = {116019},
  numpages = {10},
  year = {2022},
  month = {Dec},
  publisher = {American Physical Society},
  doi = {10.1103/PhysRevD.106.116019},
  url = {https://link.aps.org/doi/10.1103/PhysRevD.106.116019}
}

@article{hernandez2026scalar,
  author        = {Hern{\'a}ndez, A. and Zamora, R.},
  title         = {Scalar propagator in a rotating thermal medium: Consistency checks and applications},
  journal       = {arXiv},
  year          = {2026},
  eprint        = {2609.04619},
  archivePrefix = {arXiv},
  primaryClass  = {hep-ph}
}

@article{ayala2026polarization,
  author        = {Ayala, Alejandro and Medina Serna, Jos{\'e} Jorge and Dom{\'\i}nguez, Isabel and Tejeda-Yeomans, Mar{\'\i}a Elena},
  title         = {Excitation function for global $\Lambda$ polarization in relativistic heavy ion collisions with the Core Corona model},
  journal       = {arXiv},
  year          = {2026},
  eprint        = {2604.07501},
  archivePrefix = {arXiv},
  primaryClass  = {hep-ph}
}

@article{PhysRevC.105.034907,
  title = {Rise and fall of $\mathrm{\ensuremath{\Lambda}}$ and $\overline{\mathrm{\ensuremath{\Lambda}}}$ global polarization in semi-central heavy-ion collisions at HADES, NICA and RHIC energies from the core-corona model},
  author = {Ayala, Alejandro and Dom{\'\i}nguez, Isabel and Maldonado, Ivonne and Tejeda-Yeomans, Mar{\'\i}a Elena},
  journal = {Phys. Rev. C},
  volume = {105},
  issue = {3},
  pages = {034907},
  numpages = {11},
  year = {2022},
  month = {Mar},
  publisher = {American Physical Society},
  doi = {10.1103/PhysRevC.105.034907},
  url = {https://link.aps.org/doi/10.1103/PhysRevC.105.034907}
}

@article{castano2026dilepton,
  author        = {Casta{\~n}o-Yepes, J. D. and Mu{\~n}oz, E.},
  title         = {Dilepton production in a rotating thermal medium: The rigid rotation approximation},
  journal       = {Phys. Rev. D},
  volume        = {113},
  pages         = {116001},
  year          = {2026},
  eprint        = {2509.18219},
  archivePrefix = {arXiv},
  primaryClass  = {hep-ph}
}

@book{lebellac1996thermal,
  author    = {Le Bellac, Michel},
  title     = {Thermal Field Theory},
  publisher = {Cambridge University Press},
  address   = {Cambridge},
  year      = {1996},
  doi       = {10.1017/CBO9780511721700}
}

@book{kapusta2006finite,
  author    = {Kapusta, Joseph I. and Gale, Charles},
  title     = {Finite-Temperature Field Theory: Principles and Applications},
  edition   = {2},
  publisher = {Cambridge University Press},
  address   = {Cambridge},
  year      = {2006},
  doi       = {10.1017/CBO9780511535130}
}

@article{PhysRevC.93.024905,
  title = {{Centrality dependence of pion freeze-out radii in Pb-Pb collisions at $\sqrt{{s}\_{\mathit{N}N}}=2.76$ TeV}},
  author = {Adam, J. and others},
  collaboration = {ALICE Collaboration},
  journal = {Phys. Rev. C},
  volume = {93},
  issue = {2},
  pages = {024905},
  numpages = {18},
  year = {2016},
  month = {Feb},
  publisher = {American Physical Society},
  doi = {10.1103/PhysRevC.93.024905},
  url = {https://link.aps.org/doi/10.1103/PhysRevC.93.024905}
}

@article{PhysRevC.59.3324,
  title = {Size of fireballs created in high energy heavy ion collisions as inferred from Coulomb distortions of pion spectra},
  author = {Ayala, Alejandro and Jeon, Sangyong and Kapusta, Joseph},
  journal = {Phys. Rev. C},
  volume = {59},
  issue = {6},
  pages = {3324--3328},
  numpages = {0},
  year = {1999},
  month = {Jun},
  publisher = {American Physical Society},
  doi = {10.1103/PhysRevC.59.3324},
  url = {https://link.aps.org/doi/10.1103/PhysRevC.59.3324}
}

@article{PhysRevC.83.044910,
  title = {Interferometry radii in heavy-ion collisions at $\sqrt{s}=200$ GeV and 2.76 TeV},
  author = {Bo{\ifmmode \dot{z}\else \.{z}\fi}ek, Piotr},
  journal = {Phys. Rev. C},
  volume = {83},
  issue = {4},
  pages = {044910},
  numpages = {4},
  year = {2011},
  month = {Apr},
  publisher = {American Physical Society},
  doi = {10.1103/PhysRevC.83.044910},
  url = {https://link.aps.org/doi/10.1103/PhysRevC.83.044910}
}
\end{document}